\documentclass{aa}

\usepackage{blindtext}
\usepackage{graphicx}
\usepackage{multicol}
\usepackage{tabularx}
\usepackage{rotating, booktabs}
\usepackage{txfonts}
\usepackage{filecontents}
\usepackage{threeparttable}
\usepackage{hyperref}
\usepackage{natbib}
\usepackage[toc,page]{appendix}
\usepackage{subfig}
\usepackage{lscape}
\bibpunct{(}{)}{;}{a}{}{,}

\usepackage{gensymb}

\usepackage{tikz,xcolor}

\definecolor{lime}{HTML}{A6CE39}
\DeclareRobustCommand{\orcidicon}{%
        \begin{tikzpicture}
        \draw[lime, fill=lime] (0,0) 
        circle [radius=0.16] 
        node[white] {{\fontfamily{qag}\selectfont \tiny ID}};
        \draw[white, fill=white] (-0.0625,0.095) 
        circle [radius=0.007];
        \end{tikzpicture}
        \hspace{-2mm}
}

\foreach \x in {A, ..., Z}{%
        \expandafter\xdef\csname orcid\x\endcsname{\noexpand\href{https://orcid.org/\csname orcidauthor\x\endcsname}{\noexpand\orcidicon}}
}

\def\ltsima{$\; \buildrel < \over \sim \;$}
\def\simlt{\lower.5ex\hbox{\ltsima}}
\def\gtsima{$\; \buildrel > \over \sim \;$}
\def\simgt{\lower.5ex\hbox{\gtsima}}

\def\cm2{{cm$^{-2}$}}

\begin{document}

\titlerunning{Broad and narrow line Seyfert 1 galaxies}
\title{A comparative analysis of the active galactic nucleus and star formation characteristics of broad- and narrow-line Seyfert 1 galaxies}
  
\date{}

\author{{K. S. Kurian}\inst{\ref{inst1},\ref{inst2}}\orcidA{}
\and
{C. S. Stalin}\inst{\ref{inst1}}\orcidB{}
\and
{S. Rakshit}\inst{\ref{inst3}}\orcidC{}
\and
{G. Mountrichas}\inst{\ref{inst4}}\orcidD{}
\and
{D. Wylezalek}\inst{\ref{inst5}}\orcidE{}
\and
{R. Sagar}\inst{\ref{inst1}}\orcidF{}
\and
{M. Kissler-Patig}\inst{\ref{inst6}}\orcidG{}
}

\institute{Indian Institute of Astrophysics, Block II, Koramangala, Bangalore, India \email{kshama.sara@gmail.com}\label{inst1}
\and
Pondicherry University, Puducherry, 605014, India\label{inst2}
\and 
Aryabhatta Research Institute of Observational Sciences, Manora Peak, Nainital 263002, India \label{inst3}
\and
Instituto de Fisica de Cantabria (CSIC-Universidad de Cantabria), Avenida de los Castros, 39005 Santander, Spain \label{inst4}
\and
Astronomisches Rechen-Institut, Zentrum fur Astronomie der Universitat Heidelberg,
Monchhofstr. 12-14, 69120 Heidelberg, Germany\label{inst5}
\and
European Space Agency (ESA), European Space Astronomy Centre (ESAC), Camino Bajo del Castillo s/n, 28692 Villanueva de la
8 Canada, Madrid, Spain \label{inst6}
}

\abstract
{
We report here our comparative analysis of the active galactic nucleus (AGN) 
and star formation (SF) 
characteristics 
of a sample of narrow-line Seyfert 1 (NLS1) and broad-line Seyfert 1 
(BLS1) galaxies. Our sample consisted of 373 BLS1 and 240 NLS1 
galaxies and spanned the redshift 0.02 $<$ $z$ $<$ 0.8. The broad-band spectral energy 
distribution, constructed using data  
from the ultra-violet to the far-infrared, was 
modelled using CIGALE to derive 
the basic properties of our sample.  
We searched for differences in stellar mass (M$_{\ast}$), star formation 
rate (SFR), and AGN luminosity (L$_{AGN}$) in
the two populations.
We also estimated new radiation-pressure-corrected black 
hole masses for our sample of BLS1 and NLS1 galaxies.
While the virial black hole mass (M$_{BH}$) of BLS1 
galaxies is similar to their radiation-pressure-corrected M$_{BH}$ values,  the virial M$_{BH}$ values of NLS1 galaxies are underestimated. We found that 
NLS1 galaxies have a lower M$_{BH}$ of log (M$_{BH}$ [M$_{\odot}$]) = 
7.45 $\pm$ 0.27 and a higher Eddington ratio of log ($\lambda_{Edd}$)
= $-$0.72 $\pm$ 0.22 than BLS1 galaxies, which have log (M$_{BH}$[M$_{\odot}$]) and $\lambda_{Edd}$ values of 8.04 $\pm$ 0.26  and 
$-$1.08 $\pm$ 0.24, respectively.  The distributions of M$_{\ast}$, SFR, and specific star formation 
(sSFR = SFR/M$_{\ast}$) for the two populations are 
indistinguishable. This analysis is based on an independent 
approach and contradicts reports in the literature that NLS1 galaxies 
have a higher SF than BLS1 galaxies. 
While we found that L$_{AGN}$ increases with M$_{\ast}$, L$_{SF}$ flattens
at high M$_{\ast}$ for both BLS1 and NLS1 galaxies. The reason may be that SF is suppressed by AGN feedback at M$_{\ast}$ higher than $\sim$10$^{11}$ M$_{\odot}$ or that the AGN fuelling mechanism is decoupled from SF.
Separating the sample into 
radio-detected and radio-undetected subsamples, we found no difference in 
their SF properties suggesting that the effect of AGN jets on SF is negligible.}


   \maketitle

\section{Introduction}\label{sec:intro}

Narrow-line Seyfert 1 (NLS1) galaxies are a peculiar category of active galactic nuclei (AGN). Since
they were identified as a separate class of objects by \cite{1985ApJ...297..166O}, they have attracted the 
attention of the AGN community because their properties are peculiar. They have relatively
narrow permitted optical emission lines with a full width at half
maximum of the H${\beta}$ emission line $<$ 2000 km s$^{-1}$, a flux ratio
of the [OIII] to H${\beta}$ line $<$ 3, strong Fe II lines, steep soft X-ray 
spectra \citep{1996A&A...305...53B}, and rapid X-ray variability \citep{2017MNRAS.466.3309R}.
The observed properties of NLS1 galaxies are thought to be caused by 
low-mass black holes at their centres, with masses ranging from
$\sim$10$^6$ to 10$^8$ M$_{\odot}$, which  accrete at a very high 
rate \citep{2004ApJ...606L..41G,2018MNRAS.480...96W}.  
However, from spectro-polarimetric observations of the radio-loud NLS1 galaxy PKS 2004$-$447, 
\cite{2016MNRAS.458L..69B} found a polarized H${\alpha}$ line, with a width much larger than the width seen in total light, 
thereby yielding a higher black hole mass (M$_{BH}$) similar to that known for typical radio-loud AGN.  
Moreover, \cite{2013MNRAS.431..210C} and \cite{2019ApJ...881L..24V} reported based on fitting the 
accretion disk spectra to the observed spectra
of NLS1 galaxies that their $M_{BH}$ values are similar to their broad-line counterparts, namely 
broad-line Seyfert 1 (BLS1) galaxies.  They are found to show optical flux 
variations within a night \citep{2004ApJ...609...69K,2013MNRAS.428.2450P,2017MNRAS.466.2679K,2019MNRAS.483.3036O, 2022MNRAS.517.3257T}. On 
a year-like timescale, they are also found to vary in the 
optical \citep{2017ApJ...842...96R} and infrared bands \citep{2019MNRAS.483.2362R}.

The hosts of NLS1 galaxies, from observations available as of today, are
known to be spirals, often barred and with pseudo-bulges \citep{2012ApJ...754..146M,
2007ApJ...658L..13Z,2008A&A...490..583A,2014ApJ...795...58L,2016ApJ...832..157K,2018A&A...619A..69J,2020MNRAS.492.1450O,2022A&A...668A..91V}. 
A small fraction (about 5\%)  of the NLS1 galaxies is known to emit in the radio 
band, and only a handful of the sources that emit in the radio band are found to have
large-scale relativistic jets
\citep{2012ApJ...760...41D,2018ApJ...869..173R,2019MNRAS.487..640D,2022A&A...662A..20V}.  A small fraction of radio-loud NLS1 galaxies are also known to be 
emitters of 
$\gamma$-rays, as observed by the {\it Fermi} Gamma-ray Space Telescope \citep{2009ApJ...707L.142A,2019ApJ...872..169P}.
However, radio-loud objects are thought to be hosted by elliptical galaxies.
Given the uncertainties on the morphology of the 
hosts of NLS1 galaxies, it becomes
imperative to understand the hosts of these systems in relation to BLS1 galaxies.  

It has been suggested by \cite{2000MNRAS.314L..17M} that 
NLS1 galaxies are gas-rich and young sources with
ongoing star formation (SF). This also explains the high accretion and high metallicity \citep{2002ApJ...575..721N}  
in some NLS1 galaxies. From low-resolution mid-infrared spectroscopy of a very small number of 
NLS1 and BLS1 galaxies, \cite{2010MNRAS.403.1246S} found that NLS1 galaxies 
have higher SF activities than BLS1 galaxies of the same luminosity. The signatures of outflow in their spectra are also stronger than in BLS1 galaxies
\citep{2022MNRAS.510.4379J}. Given the evidence of outflows
and considering that NLS1 are galaxies with a high Eddington ratio that may produce high radiative feedback,
it is also interesting to determine whether the AGN and 
the SF characteristics of the host galaxies are connected.

The cosmic SF rate and black hole accretion rate follow a similar 
evolution over cosmic time. Both of them show a peak at around 
$z$ $\sim$2 
that is followed by a sharp decline towards the present age \citep{2014ARA&A..52..415M,
2003ApJ...587...25D, 2007A&A...474..755B,2007ApJ...654..731H}. 
Observational evidence also indicates a close correlation between (a) the mass 
of the central super-massive black hole and the galaxy luminosity 
\citep{1995ARA&A..33..581K,2003ApJ...589L..21M,2009ApJ...698..198G}, (b) mass of 
the super-massive black hole and the galaxy bulge mass \citep{1998AJ....115.2285M,2002MNRAS.331..795M}, 
and (c) the mass of the super-massive black hole and the velocity dispersion \citep{2000ApJ...539L...9F,
2000ApJ...539L..13G,2001ApJ...547..140M}. All available observations thus 
indicate that black holes (their formation and growth) and their host galaxy 
properties are fundamentally coupled. Studies are also available in the literature
about the connection between AGN and their host galaxies, but the results
of these studies disagree. For example, there 
are reports that the AGN luminosity (L$_{AGN}$) correlates with the star formation rate (SFR) of their host
galaxies \citep{2017A&A...602A.123L,2020ApJ...896..108Z}, while some studies have found either a weak or no correlation
between SFR and L$_{AGN}$ \citep{2012ApJ...760L..15H,2017MNRAS.466.3161S,2017MNRAS.472.2221S}. 
The conflicting results from different studies might in part be due to 
the sample used for the studies and the analysis methods that were followed. 

A close correlation is known to exist between SF and stellar
mass (M$_{\ast}$) for SF galaxies of the main 
sequence (MS;\citealt{2007ApJ...670..156D,2007A&A...468...33E}). In the case of AGN, a few studies
indicated that the host galaxies of AGN lie on or above the MS 
\citep{2009ApJ...696..396S, 2012A&A...540A.109S,2022ApJ...934..130Z}, 
while other studies reported evidence that AGN host galaxies lie below
the MS. 
\citep{2012MNRAS.427.3103B,2015MNRAS.453L..83M,2015MNRAS.452.1841S}. 
One of the reasons for these results that are known
today might be that the AGN activity plays an important role in
regulating the SF in their host galaxies via the feedback processes
that operate in them. Thus, the effect of the AGN on their host
galaxies is still a matter of debate, and a systematic investigation of the
AGN activity and its impact on the SF activity in their hosts
is indeed needed to assess the nature of the connection between AGN and 
SF activity.

We aim to investigate the nature of the host galaxies of NLS1 galaxies and determine how they 
compare with the host galaxy properties of a comparison sample of BLS1 galaxies. This investigation will also enable us to understand
the effect of the central AGN in NLS1 and BLS1 galaxies has on their hosts in the context
of them (as is generally thought) being powered by low- and high-mass black holes, respectively. The main controversies surrounding NLS1 galaxies and their association with BLS1 galaxies are (1) the differences in the black hole mass, with NLS1 galaxies thought to have a lower black hole mass than BLS1 galaxies, (2) the high Eddington ratios of NLS1 with respect to BLS1 galaxies, and (3) the claim that NLS1 have higher SFRs and are hosted by young gas-rich galaxies with respect to BLS1 galaxies. Although many comparative studies of NLS1 and BLS1 are published, their samples have included very few objects, or they only compared one of the galaxy properties. In this work, we therefore carried out a comparative analysis of the physical parameters of the AGN and the host galaxies in populations of NLS1 and BLS1 galaxies. The paper is 
organised as follows. In Section 2 we describe the selection of the sample and
the data collection. 
The broad-band spectral energy distribution (SED) fitting is described in 
Section 3. In section 4 we investigate the similarities and/or differences of the 
host galaxy properties of NLS1 and BLS1 galaxies, and in the final section, we summarise our
results.

\section{Sample and data}
Our initial sample of NLS1 galaxies was taken from \cite{2017ApJS..229...39R}, 
while the BLS1 galaxies were taken from those obtained in the process of arriving at 
the NLS1 galaxy sample by \cite{2017ApJS..229...39R}. They span the redshift 
0.02 $<$ $z$ $<$ 0.8. The  sample consisted of 11101 and 14886 NLS1 and BLS1 
galaxies. According to \cite{2009MNRAS.399.1206H}, the far-infrared (FIR) emission
in AGN host galaxies mostly arises from SF. To 
characterise the SF properties of galaxies hosting AGN, 
the inclusion of FIR data is therefore indeed important because it can constrain the AGN 
contribution to the infrared luminosity of the host galaxy, and it improves the SFR 
estimation without AGN contamination \citep{2018MNRAS.478.3721S}. When FIR photometry is not included, the SFR is known to be  
systematically underestimated \citep{2018A&A...618A..31M}. Therefore, we selected all sources that were detected by the {\it Herschel} Space Observatory in 
at least one band of 100, 160, 250, 350, and 500 $\mu$m. With
the FIR constraint, our sample for this study consisted of 
240 NLS1 galaxies and 373 BLS1 galaxies. Based on the available multi-wavelength 
data set, this is the best sample that can be constructed for a comparative 
analysis of the SF characteristics of BLS1 and NLS1 galaxies. 
This supersedes the single earlier study  \citep{2010MNRAS.403.1246S} with a
manifold increase in the number of sources, and we use a different approach. The FIR selection criteria bias the sample towards high SFR galaxies. However, this bias affects both BLS1 and NLS1 galaxies similarly, and hence, it does not affect the results of a comparative analysis of the two types of galaxies. The observed fluxes and their corresponding errors for our sample of BLS1 galaxies and NLS1 galaxies are given in the Appendix in Tables \ref{table-4} and \ref{table-5}, respectively. We note that the selected sample of 
240 NLS1 and 373 BLS1 galaxies was for a broad-band
SED modelling, but the final sample selected for the comparative
study was based on an SED fitting (see Section 3).

We collected the broad-band photometric data for all our sample sources 
based on observations carried out by various ground- and space-based telescopes. In the 
ultra-violet (UV) region, we used data in the near-UV (NUV) and far-UV (FUV) bands 
from the Galaxy Evolution Explorer (GALEX;\citealt{2007ApJS..173..682M}) archives. 
In the optical band, we used data in ugriz photometric bands from the 
Sloan Digital Sky Survey (SDSS\footnote{https://www.sdss4.org/dr17/}). Near 
infra-red (NIR) data in J, H, and K bands were taken from the Two Micron All-Sky 
Survey (2MASS;\citealt{2006AJ....131.1163S}).  For the mid-IR, we used data in the 
W1, W2, W3, and W4 bands from the {\it Wide-field Infrared Survey Explorer 
(WISE};\citealt{2010AJ....140.1868W}), and for the FIR, we used data from the 
Herschel Space Observatory ({\it Herschel}; \citealt{2010A&A...518L...1P}).
The photometric data  and their associated errors used in this work thus come 
from different instruments
with various point spread functions and methods adopted to derive the 
magnitudes. Both BLS1 and NLS1 galaxies are also known to 
show flux variations, and therefore, the non-simultaneous measurements
collected across different wavelengths affect the SED modelling. However, 
as both the BLS1 and NLS1 used in this work suffer from the same issues of (a) the 
non-uniformity of the aperture sizes used for the magnitude determinations and
(b) the non-simultaneity of flux measurements, their 
impact, if any, on a comparative analysis of the derived physical quantities may not 
be significant.

\begin{table*}
\setlength{\tabcolsep}{0.04pt}

\centering
\caption{Parameters adopted for the SED fitting using CIGALE}\label{table-1}
\begin{tabular}{ccc}
\hline
\hline
Parameter & Description & Value \\
\hline
& SFH - Delayed Model & \\
\\
Age\_main &     Stellar age     &   200, 500, 700, 1000, 2000                 \\
                               &&  3000, 4000, 5000 Myr                                              \\  
$\tau\_main$  & e-folding time of the stellar population & 1500, 2000, 3000, 4000, 5000,        \\
              &                                               & 7000, 10000, 12000 Myr               \\
Age\_burst, $\tau\_burst$    & Age and e-folding time of the late burst                         & 10000 Myr, 50 Myr                            \\
F\_burst      & Mass fraction of the late burst population    & 0.0, 0.005, 0.01, 0.015, 0.02, 0.05, \\
              &                                               & 0.10, 0.15, 0.18, 0.20               \\

\hline
 &  Stellar Emission  \cite{2003MNRAS.344.1000B}&\\
 \\
IMF            &  Initial mass function     & Chabrier \\  
Z              &  abundance               & 0.02  \\
\hline

              &  Dust attenuation \cite{2000ApJ...539..718C}    &\\
              \\
Av\_ISM           & V band extinction in the ISM  &  0.2, 0.3, 0.4, 0.5, 0.6, 0.7, 0.8, 0.9, 1.0, \\
                  &                               &  1.5, 2.0, 2.5, 3.0, 3.5, 4.0 \\

\hline
& Nebular Emission &\\
\\
log U         & logarithm of the Ionization parameter & -2.0  \\
Line\_width   & Width of the line in km/s                              & 300.0 \\
\hline
              & Dust \cite{2014ApJ...784...83D} &\\
              \\
$\alpha$  & power law index of the sum of the dust templates & 2.0  \\
frac\_AGN &  AGN fraction                                                                 & 0.0 \\
\hline
             & AGN module {\it skirtor2016:} \cite{2016MNRAS.458.2288S} & \\
             \\
i      & viewing angle                                                         & 30, 70 \\
Delta  & power law index of the optical slope               & -0.36 \\
f$\rm_{AGN}$  & AGN fraction & 0.0, 0.1, 0.2, 0.3, 0.4, 0.5,  \\
              &              & 0.6, 0.7, 0.8, 0.9, 0.99                \\
E(B-V)        & Extinction in the polar direction    & 0.0, 0.2, 0.4 \\
\hline
                     
\end{tabular}
\end{table*}

\section{Spectral energy distribution modelling using CIGALE}


We derived various parameters of the host galaxies of our sample using
the Code Investigating GALaxy Emission (CIGALE; \cite{2019A&A...622A.103B}). CIGALE
is a spectral energy distribution (SED) modelling code that relies on the availability of photometric data in
multiple wavelengths to derive the properties of galaxies by comparing the modelled
SEDs to observed SEDs. CIGALE works on the principle
of energy balance (i.e. the energy emitted in the mid- and far-IR
bands by dust matches the energy absorbed by dust in the visible and UV
bands) and uses a Bayesian analysis method to derive the model parameters \citep{2019A&A...622A.103B}.
The fitting routine adopted here takes into account (a) the radiation emitted
by stars that dominate the optical region from 3000 to 4000 \AA,  ~(b) the
radiation from dust heated by stellar emission that dominates the
FIR region, and (c) the radiation from the accretion disk in AGN, which peaks in
the UV region, as well as the scattered radiation
by the dusty torus, which peaks in the mid-IR region. We briefly describe
the modules we adopted for the SED fitting below.  

\begin{enumerate}
\item {\it sfhdelayed module:} To
   model the SEDs, we used the delayed star formation history (SFH), which is
   expressed as
   
\begin{equation}
SFR(\tau) \propto \frac {t}{\tau^2} * exp (\frac {-t}{\tau}) \qquad for ~0 < t < t_o.
\end{equation}


Here, $\tau$ is the time
at which the SFR peaks, and $t_o$ is the age of the onset of SF. In this model, the SFR decreases smoothly after
peaking at t = $\tau$. Using simulated realistic SEDs of galaxies with AGN 
and adopting various SFH models in the CIGALE modelling of the simulated 
SEDs, \cite{2015A&A...576A..10C} have shown that the {\it sfhdelayed} model 
provides reliable M* and SFR measurements and should be preferred
over the single exponentially decreasing and double exponentially decreasing
models. Moreover, \cite{2022A&A...663A.130M} recently compared 
the {\it sfhdelayed} with the newest  {\it sfhdelayedbq} available in CIGALE, which 
allows both an instantaneous recent variation of the SFR upwards (burst) and 
downwards (quenching; \citealt{2017A&A...608A..41C,2019A&A...622A.103B}). The authors 
found that the two SFH models provide a consistent SFR.  We 
therefore adopted the {\it sfhdelayed model} in this work.

\item {\it BC03 module:} To generate the spectrum for the adopted SFH, we adopted
the single stellar population library of \cite{2003MNRAS.344.1000B} for a
\cite{2003PASP..115..763C} initial mass function (IMF) and solar metallicity. We adopted a value of
10 Myr as the age separating the old from the young stellar
populations.

\item {\it dustatt\_modified\_CF00 module:} The dust present in galaxies absorbs UV to optical radiation
and re-emits them in the mid- and far-IR bands. To model the attenuation of
starlight by the dust in galaxies,
we used the {\it dustatt\_modified\_CF00} module, which
implements the model of \cite{2000ApJ...539..718C}. This model
assumes two power-law attenuation curves of the form
$A(\lambda) \propto \lambda^{\alpha}$, one for the birth cloud (BC),
and the other for the interstellar medium (ISM). For the attenuation due to the
ISM and BC, we assumed a power-law index $\alpha$ of $-$0.7.

\item {\it \cite{2014ApJ...784...83D} module:} To model the IR emission from the dust heated by radiation from stars, we used
the templates from \cite{2014ApJ...784...83D}. This also comprises the radiation from dust heated by the AGN. As the AGN model is
included separately, we set the AGN contribution in
\cite{2014ApJ...784...83D}  model to 0. This model is parametrised as  
\begin{equation}
dM_{dust}(U) \propto U^{-\alpha} dU.
\end{equation}
Here, U is the
radiation field intensity, and M$_{dust}$ is the mass of the dust heated by the radiation field. We chose $\alpha$ to have a range of
values as given in Table \ref{table-1}.

\item {\it AGN module:} Along with the SF, the AGN in the hosts contribute to the
observed emission from galaxies, and it is difficult to distinguish this because the AGN and stars both emit in the UV band, and
a large fraction of this UV emission is absorbed by dust in galaxies
and is re-emitted at mid- and far-IR wavelengths. To model the contribution of AGN
to the observed emission from galaxies, we used the SKIRTOR
\citep{2016MNRAS.458.2288S} model to parametrise the AGN component to the
SED. This is based on SKIRT, a 3D radiative transfer code that includes obscuration by the torus \citep{2011ApJS..196...22B}.
\end{enumerate}

\begin{figure*}
      \hbox{
           \hspace*{-0.5cm}\includegraphics[scale=0.57]{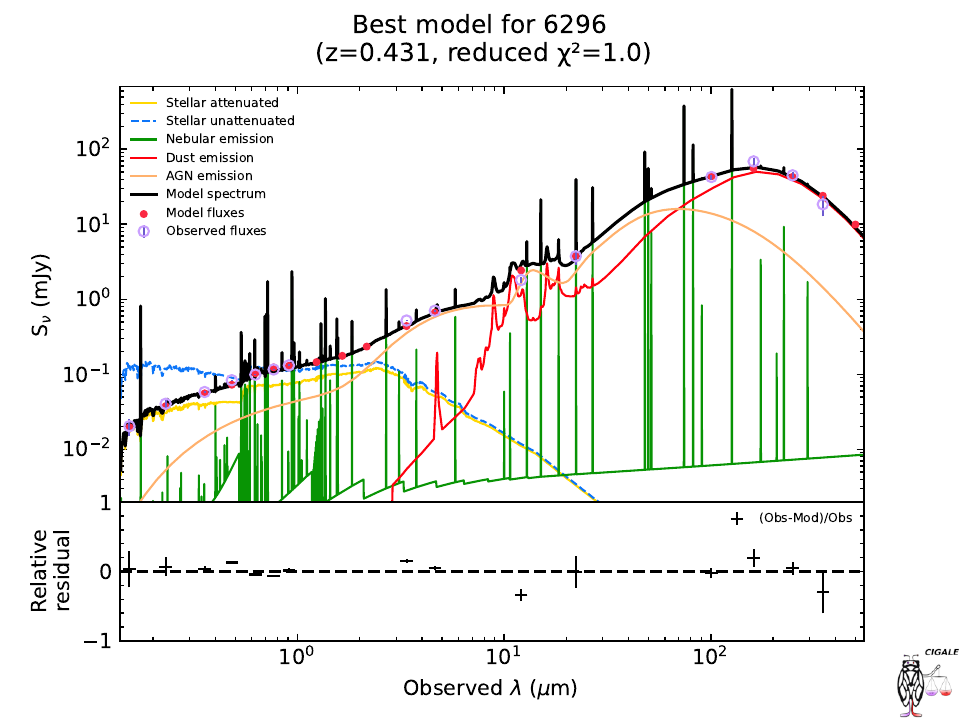}
           \includegraphics[scale=0.57]{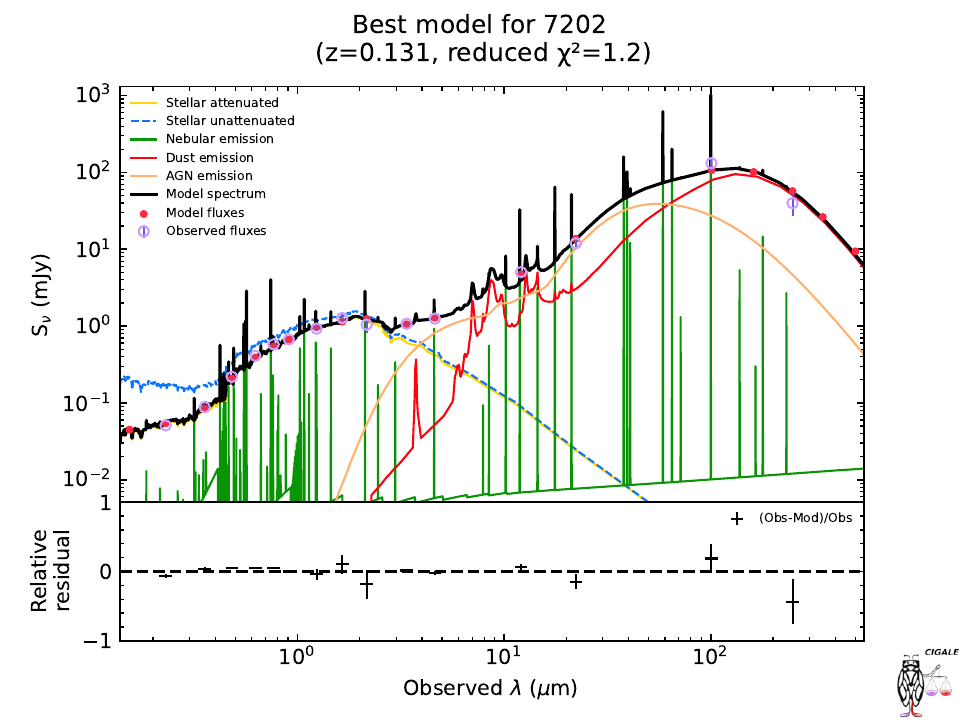}
           }
\caption{Examples of the observed SED along with the best-fit SED from CIGALE for one NLS1 galaxy (left panel) and one BLS1 galaxy (right panel).
The open and filled symbols correspond to the observed and modelled flux densities. The goodness of the fit, represented by the reduced
$\chi^2$, is given at the top of each plot. The residuals of the fit are given at the bottom of each panel.}
\label{figure-1}
\end{figure*}

\begin{figure}
\vbox{
\includegraphics[scale=0.55]{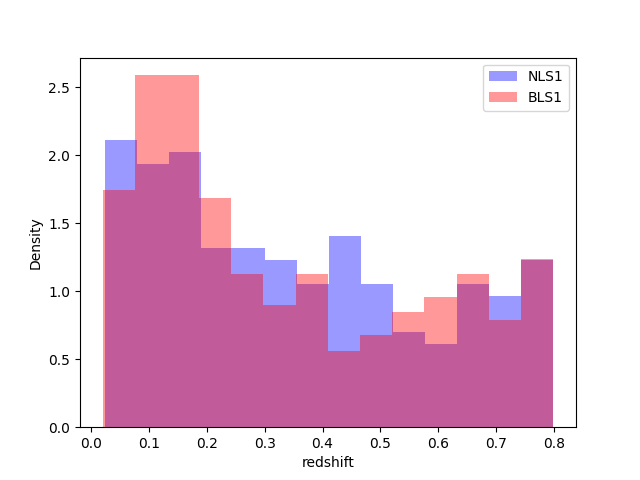}
}
\caption{Redshift distribution for the best SED fits of the NLS1 and BLS1 galaxies}
\label{figure-2}
\end{figure}

The complete set of parameters we used to build the SEDs of our sample of galaxies is
given in Table \ref{table-1}.
The parameters we adopted for the SED modelling are similar to those used by 
\cite{2021A&A...653A..70M,2022A&A...659A.129V,2022A&A...658A..35K} and were modified 
to suit our sample of sources. The physical parameters from this SED
analysis were derived through the analysis of the likelihood distribution. Examples of SED fits
to the data are given in Fig. \ref{figure-1}.
We note that an SED analysis of this type can have errors due to (a) the 
different aperture sizes used in the photometric
measurements taken from different instruments, and (b) the photometric 
measurements being not simultaneous. Our sample sources are known to vary, which may
lead to errors in the model fitting. 
The median $\chi^2$ per
degrees of freedom for the BLS1 sample is 2.5. Therefore, for all subsequent analyses,
we used only the sources for which the SED fitting returned a $\chi^2$ per 
degrees of freedom values ranging
from 0.5 (a conservative limit) to 5 (twice the median value of BLS1). With these criteria, we were left with a final sample
of 319 BLS1 and 205 NLS1 galaxies. All further analyses were restricted to 
this
reduced sample. These two samples match in redshift, as shown in Fig. \ref{figure-2}.
The mean $z$-values of this reduced sample are
0.33 and 0.35 for BLS1 and NLS1 galaxies, respectively. A KS test 
$p$-value of 0.14 shows that the redshift distribution of both 
types of galaxies is similar. From previous studies using CIGALE, it is known that the uncertainties in CIGALE are underestimated, and the true uncertainties in SFR and stellar mass are at least 0.3 dex (\cite{2019MNRAS.485.2710J} and the references therein). Some of the physical parameters obtained 
from CIGALE modelling of the observed SEDs are given in Table \ref{table-2} and 
Table \ref{table-3} for theBLS1 and NLS1 galaxies, respectively.

\begin{table*}
\caption{Parameters derived for the sample of BLS1 galaxies by fitting the 
SED with CIGALE. The table in full is available in the electronic version of the article.}
\label{table-2}
\scalebox{0.75}{
\begin{tabular}{lcccccccccccccc}
\hline
\\
 index & SDSS ID & RA & Dec & z & FWHM & L$_{5100}$ & L$_{AGN}$ & log$_{10}$(SFR) & L$_{SF}$ & M$_{\ast}$ & log$_{10}$(sSFR) & M$_{BH}$ & M$_{BH}$ & log$_{10}$($\lambda_{Edd}$) \\
    &  &  &  &  &  & & &  &  &  &  & (virial) & (RP) & \\
   &  & deg & deg &  & km s$^{-1}$ & erg s$^{-1}$ & erg s$^{-1}$ & M${_\odot}$ yr$^{-1}$ & erg s$^{-1}$ & M${_\odot}$ & Gyr$^{-1}$ & M${_\odot}$ & M${_\odot}$ & \\
 \\\hline
\\
1 &1768-53442-0193 &187.7193 &14.5515 &0.1151 &3649.26 &43.49 &43.68 &0.68 &44.22 &10.37 &-0.69 &7.54 &7.42 &-1.85 \\

2 &7235-56603-0309 &37.2306 &-5.1918 &0.3656 &2451.56 &44.06 &44.82 &1.6 &45.14 &11.06 &-0.47 &7.5 &7.84 &-1.13  \\

3 &0470-51929-0431 &135.589 &0.5725 &0.3262 &3568.51 &44.24 &45.08 &1.73 &45.27 &10.91 &-0.18 &7.92 &8.06 &-1.09\\

4 &1616-53169-0453 &189.39 &13.3185 &0.1511 &2464.65 &43.67 &44.37 &0.94 &44.48 &10.35 &-0.42 &7.3 &7.48 &-1.23 \\

5 &0471-51924-0172 &136.9312 &1.5578 &0.1642 &4905.72 &43.71 &44.28 &-0.51 &43.04 &11.28 &-2.78 &7.92 &7.7 &-1.53 \\
\\
\hline
\end{tabular}
}
\end{table*}

\begin{table*}
\caption{Parameters derived for the sample of NLS1 galaxies by fitting the 
SED with CIGALE. The table in full is available in the electronic version of the article.}
\label{table-3}
\scalebox{0.75}{
\begin{tabular}{lcccccccccccccc}
\hline
\\
 index & SDSS ID & RA & Dec & z & FWHM & L$_{5100}$ & L$_{AGN}$ & log$_{10}$(SFR) & L$_{SF}$ & M$_{\ast}$ & log$_{10}$(sSFR) & M$_{BH}$ & M$_{BH}$ & log$_{10}$($\lambda_{Edd}$) \\
    &  &  &  &  &  & & &  &  &  &  & (virial) & (RP) & \\
   &  & deg & deg &  & km s$^{-1}$ & erg s$^{-1}$ & erg s$^{-1}$ & M${_\odot}$ yr$^{-1}$ & erg s$^{-1}$ & M${_\odot}$ & Gyr$^{-1}$ & M${_\odot}$ & M${_\odot}$ & \\
 \\\hline
 \\
1 &7386-56769-0752 &156.753 &48.5549 &0.5985 &1512 &43.46 &44.74 &0.56 &44.1 &10.44 &-0.88 &6.76 &7.23 &-0.6\\

2 &2751-54243-0611 &226.5035 &14.2612 &0.1432 &1950 &44.21 &45.25 &1.62 &45.16 &10.38 &0.25 &7.38 &7.96 &-0.82 \\

3 &1959-53440-0473 &157.1984 &31.7738 &0.2067 &1364 &43.76 &45.02 &1.3 &44.84 &10.71 &-0.4 &6.83 &7.51 &-0.6 \\

4 &2216-53795-0107 &171.5435 &26.6513 &0.2949 &2039 &43.5 &45.36 &1.64 &45.18 &10.29 &0.35 &7.04 &7.3 &-0.05 \\

5 &6445-56366-0710 &163.0503 &32.2317 &0.7843 &1194 &43.62 &44.38 &1.14 &44.68 &10.44 &-0.3 &6.64 &7.36 &-1.1\\

\\
\hline
\end{tabular}
}
\end{table*}

\begin{figure*}
\vbox{
\includegraphics[width=1.0\columnwidth]{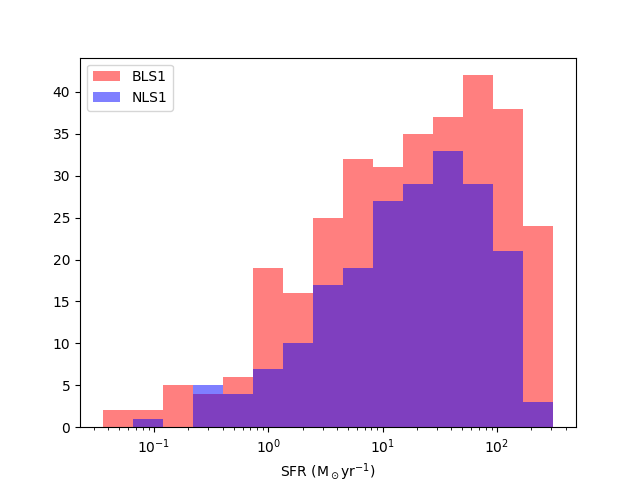}
\includegraphics[width=1.0\columnwidth]{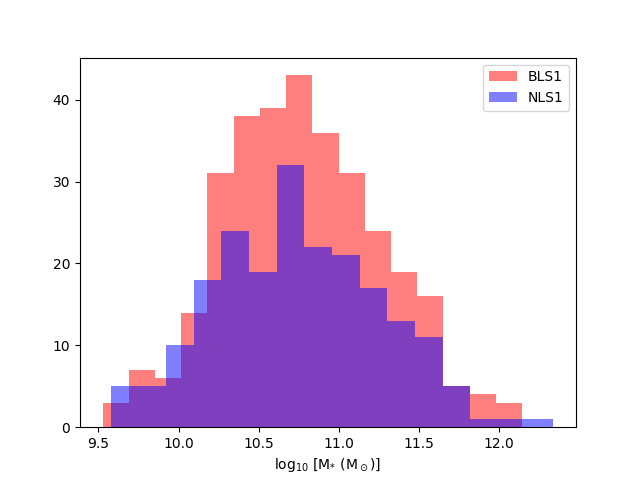}
     }
\caption{Histogram of the SFR (left) and M$_{\ast}$ (right) 
for our sample of BLS1 and NLS1 galaxies.
}
\label{figure-3}
\end{figure*}

\begin{figure*}
\vbox{
\includegraphics[width=1.0\columnwidth]{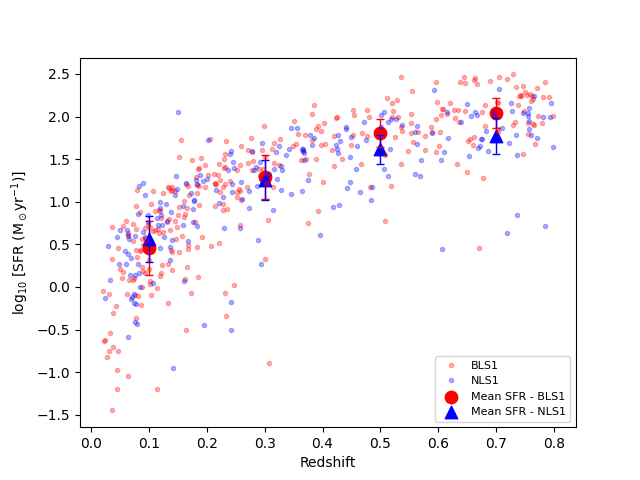}
\includegraphics[width=1.0\columnwidth]{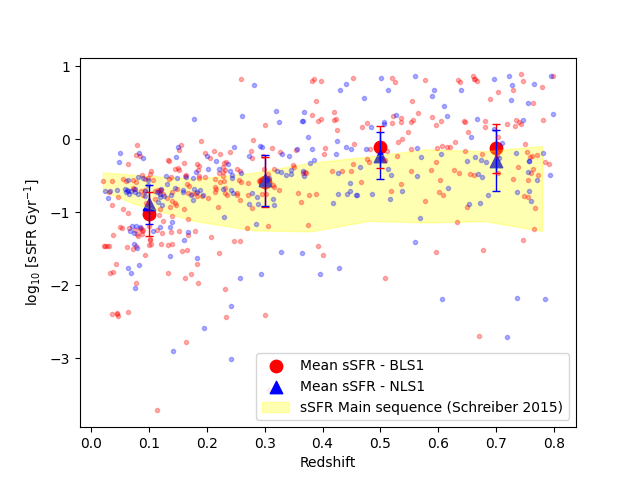}
    }
\caption{Star formation properties of Seyfert 1 galaxies. Left panel: Redshift-binned mean values of SFR of BLS1 (red) and 
NLS1 (blue) galaxies. Right panel:  Redshift-binned sSFR of  BLS1 (red) and 
NLS1 (blue) galaxies. The shaded yellow region represents the sSFR of MS galaxies for the stellar mass range of the BLS1 and NLS1 galaxies.
}
\label{figure-4}
\end{figure*}

\begin{figure*}
\vbox{
\includegraphics[width=1.0\columnwidth]{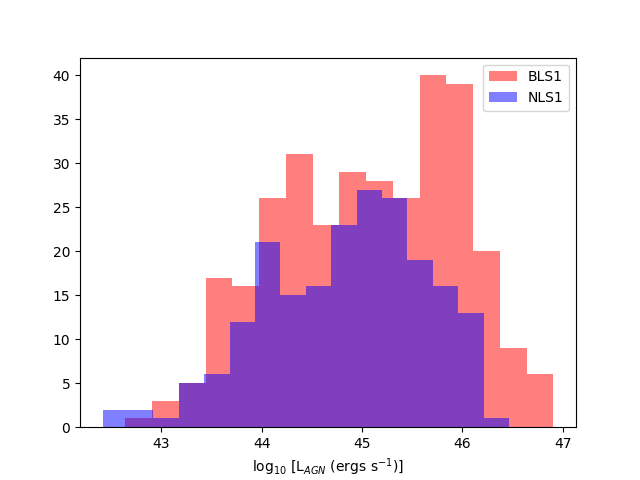}
\includegraphics[width=1.0\columnwidth]{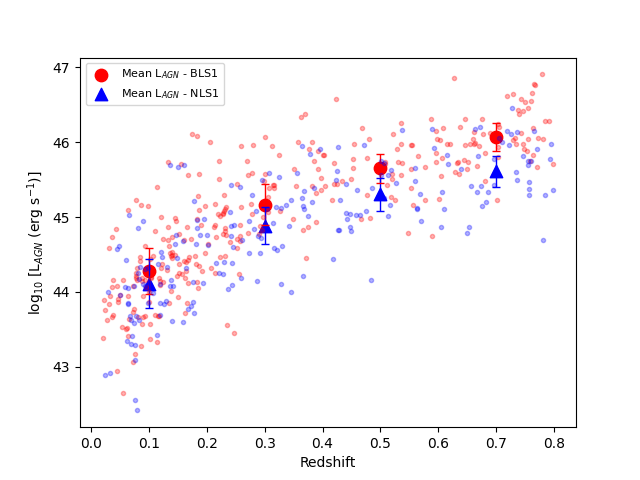}
    }
\caption{AGN luminosity of Seyfert 1 galaxies. Left panel: Distribution of L$_{AGN}$ for 
BLS1 and NLS1 galaxies. Right panel: Variation in L$_{AGN}$ with
redshift.  
}
\label{figure-5}
\end{figure*}

\section{Results}

In this section, we compare the AGN and SF properties of BLS1 and 
NLS1 galaxies and show that the AGN activity might affect the SF in the host galaxy.

\subsection{Host galaxy properties: Star formation rate, stellar mass, and specific star formation rate}

We show in Fig. \ref{figure-3} the distribution of the SFR and M$_{\ast}$ 
 for our sample of BLS1 and NLS1 galaxies. For the NLS1 galaxies,
we found a mean log(SFR M${_\odot}$yr$^{-1}$) = 1.18 $\pm$ 0.33, while for the BLS1 galaxies, we found 
a mean log(SFR M${_\odot}$yr$^{-1}$) = 1.18 $\pm$ 0.41. We thus found no difference in the SFR
between NLS1 and BLS1 galaxies. A KS test indicates that the distributions are
similar at the 90\% level, with a $p$-value of 0.07. The distribution of 
M$_\ast$ for our sample of NLS1 and BLS1 galaxies is shown in the right 
panel of 
Fig. \ref{figure-3}. We found a mean M$_\ast$ of log (M$_\ast$[M${_\odot}$]) = 
10.74 $\pm$ 0.25
and 10.78 $\pm$ 0.24 for the NLS1 and BLS1 galaxies, respectively. From a KS test, we found 
that the distribution of M$_\ast$ is similar for the BLS1 and NLS1 galaxies, 
with a $p$-value of 0.71.

We show in Fig. \ref{figure-4} the evolution of thge SFR and sSFR (SFR/M$_{\ast}$)
with $z$. The SFR and sSFR are binned in redshift, and the geometric mean of the SFR and sSFR in each redshift bin is plotted over the individual data points. We also computed the SFR of main sequence (MS) galaxies (SFR$_{MS}$) 
using the relation described in \cite{2015A&A...575A..74S},
\begin{equation}
log(SFR_{MS}) = m - m_0 + a_0*r - a1[max(0,m - m_1- a_2r)]^2 .
\end{equation}

Here, $m_0$ = 0.5 $\pm$ 0.07, $a_0$ = 1.5 $\pm$ 0.15, $a_1$ = 0.3 $\pm$ 0.008, 
$m_1$ = 0.36 $\pm$ 0.3, $a_2$ = 2.5 $\pm$ 0.6, r = log$_{10}$(1+z), and m = $log_{10}(M_{\ast}/M_{\odot}$).
The generalised relation arrived at by \cite{2015A&A...575A..74S}  was (a) 
based on objects that span a range of redshifts from 0.3 $<$ $z$ $<$ 5.0, and (b) the 
SFR was estimated by combining direct UV and reprocessed UV light in the FIR. 
The redshifts of our sample sources overlap with the range investigated 
by \cite{2015A&A...575A..74S} and also use both UV to FIR photometric points 
to obtain the SFR. {\bf The \cite{2015A&A...575A..74S} MS relation extends down to $z=0$ (see \citealt{2017A&A...608A..41C}).}  We therefore adopted the relation of \cite{2015A&A...575A..74S}. 
Analytical expressions from the literature for the MS may 
introduce systematic effects. These effects are due to the different methods 
that were applied to calculate the host galaxy properties, for instance, to the different models 
and parametric grid for the SED fitting, to the different photometric coverage, 
the different selection criteria, and even to the different definitions of the MS 
\citep{2021A&A...653A..74M}. \cite{2021A&A...653A..74M} compared in their Fig. 6 the 
SFR calculations of CIGALE (using sfhdelayed) with the SFR from \cite{2015A&A...575A..74S} 
and \cite{2014ApJ...795..104W}. They found that at low redshifts ($z<$1.5), the 
SFR from CIGALE tends to be lower than the SFR from \cite{2015A&A...575A..74S}. 
However, this difference is about 0.25 dex and does not affect the results and 
conclusions of this paper (e.g. the right panel in \ref{figure-4}).
The shaded yellow region in Fig. \ref{figure-4} represents the sSFR of MS galaxies as described in Equation 3 
for the stellar mass range of BLS1 and
NLS1 galaxies. The mean log(sSFR) of NLS1 and BLS1 galaxies
tends to lie on the SF main sequence in the sSFR-$z$ plane. This indicates
that the SF properties of Seyfert 1 galaxies are similar
to the SF properties of MS galaxies, and this agrees
with studies that showed AGN to be preferentially hosted in star-forming galaxies \citep{2012ApJ...753L..30M}.

\subsection{Active galactic nucleus properties: Active galactic nucleus luminosity, black hole mass, and Eddington ratio}

\begin{figure}
\vbox{
\includegraphics[scale=0.75]{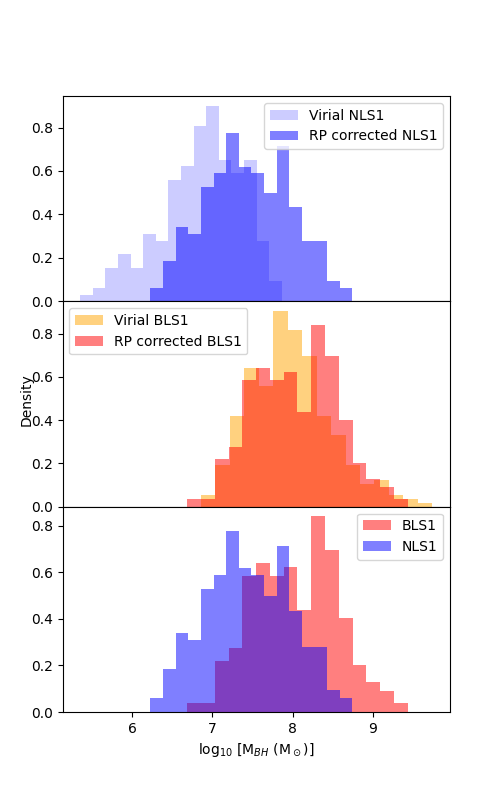}
     }
\caption{Distribution of M$_{BH}$ values of BLS1 and NLS1 galaxies. 
\textit{Top panel:} Virial (light blue) and radiation-pressure-corrected 
(dark blue) M$_{BH}$ values of NLS1 galaxies. \textit{Middle panel:} Virial 
(orange) and radiation-pressure-corrected (red) M$_{BH}$ values of BLS1 
galaxies. \textit{Bottom:} Radiation-pressure-corrected M$_{BH}$ values 
of NLS1 (blue) and BLS1 (red) galaxies.}
\label{figure-6}
\end{figure}

To model the AGN emission, we used the skirtor2016 module in CIGALE (see Table 1). 
SKIRTOR is a clumpy two-phase torus model \citep{2016MNRAS.458.2288S} and is based 
on SKIRT \citep{2011ApJS..196...22B}, which is a 3D radiative transfer 
code\footnote{https://skirt.ugent.be/root/index.html}. In this module, 
the parameter $f_{AGN}$ defines the relative strength between AGN and galaxy 
components, and we varied $f_{AGN}$ from 0.0 to 0.99. $L_{AGN}$ is one of 
the output parameters of the skirtor2016 
module and is the sum of the AGN disk luminosity and AGN dust luminosity.
We show in the left panel in Fig. \ref{figure-5} the distribution
L$_{AGN}$ for our sample of BLS1 and NLS1 galaxies.
We found mean log (L$_{AGN}$) values
of 45.06 $\pm$ 0.44 and 44.84 $\pm$ 0.39 for BLS1 and NLS1 galaxies, respectively. This shows that BLS1 galaxies
tend to have L$_{AGN}$ slightly higher than NLS1 galaxies. The KS test also
confirms that the distribution of L$_{AGN}$ for BLS1 and NLS1 galaxies
is different, with a $p$-value of 0.001. When binned in $z$, the NLS1 galaxies
have low L$_{AGN}$ than the BLS1 galaxies (see Fig. \ref{figure-5} 
right panel).

To estimate the Eddington ratio, we need robust estimates of black hole 
masses.  It is generally thought that NLS1 galaxies are powered by low-
mass black holes \citep{2004ApJ...606L..41G,2018MNRAS.480...96W} compared to 
BLS1 galaxies (but see also \cite{2013MNRAS.431..210C,2016MNRAS.458L..69B,2019ApJ...881L..24V} for an alternative view). According 
to \cite{2008ApJ...678..693M}, while determining M$_{BH}$ from the use of 
virial theorem to the broad emission lines in 
AGN spectra, the effect of radiation 
pressure needs to be taken into account. When the radiation pressure is not included, the M$_{BH}$ values in sources radiating close
to the Eddington limit would be underestimated. \cite{2008ApJ...678..693M} arrived at an empirical 
relation to determine M$_{BH}$ that also considers the effect of 
radiation pressure. We therefore recalculated the M$_{BH}$ values for the sources 
in our sample of BLS1 and NLS1 galaxies using the relation given in 
\cite{2008ApJ...678..693M} ,
\begin{equation}
\frac{M_{BH}} {M_{\odot}} = \tilde{f}\left(\frac{V_{H\beta}} {1000 \, km \, s^{-1}}\right)^{2}\left(\frac{L_{5100}} {10^{44} \, ergs \, s^{-1}}\right)^{0.5}+\tilde{g}\left(\frac{L_{5100}} {10^{44} \, ergs \, s^{-1}}\right).
\end{equation}
Here, V$_{H\beta}$ is the full width at half maximum of the H$\beta$ 
emission line, $L_{5100}$ represents the luminosity at 5100 $\AA$, and $\tilde{f}$ and $\tilde{g}$ are the scale factors that take
the physical and geometrical properties of the broad-line region into account. 
The parameter $g$ depends on the cloud mass via the assumed column density
and sets the relative importance of gravity and radiation pressure. We used a log ($\tilde{f}$) value of 6.13 and 
log ($\tilde{g}$) value of 7.72, following \cite{2008ApJ...678..693M}. 

For the virial M$_{BH}$, we used the values given in \cite{2017ApJS..229...39R}, which were estimated as
follows:

\begin{equation}
M_{B} = f R_{BLR} \Delta v^2/G.
\end{equation}
Here, $\Delta$v is the full width at half maximum of the broad emission line, and $R_{BLR}$ is the
radius of the broad-line region, determined as
\begin{equation}
log\left(\frac{R_{BLR}}{lt - day}\right) = K + \alpha \times log \left(\frac{\lambda L_{\lambda} (5100 \AA}{10^{44} erg s^{-1}}\right),
\end{equation}
where the values of K and $\alpha$ are 1.527 and 0.533, respectively, taken from \cite{2013ApJ...767..149B}.
Considering a spherical distribution of clouds, the scale factor $f$ = 3/4 \citep{2017ApJS..229...39R}. 
We show in Fig. \ref{figure-6} the distribution of the virial and pressure-corrected M$_{BH}$ values for both BLS1 and NLS1 galaxies in the
top two panels. For the BLS1 galaxies, the distribution of 
the virial and pressure-corrected M$_{BH}$ values is nearly similar, with a
mean log(M$_{BH}$[M$_{\odot}$) of 7.98 $\pm$ 0.25 and 8.04 $\pm$ 0.26, 
respectively. However, the KS test rejects the null hypothesis that 
the distributions are indeed drawn from the same population with a
$p$-value of 0.001.
For the NLS1 galaxies, 
the distributions of the virial  M$_{BH}$ values slightly overlap
the distribution of the pressure-corrected  M$_{BH}$ values, but
both distributions are indeed different based on the KS test statistics 
($p$ = 8.7 $\times$ 10$^{-16}$).  In the bottom panel of Fig. \ref{figure-6}, we show the distribution
of the pressure-corrected M$_{BH}$ values for both NLS1 (mean 
log(M$_{BH}$[M$_{\odot}$) = 7.45 $\pm$ 0.27)  and BLS1 galaxies 
(mean log(M$_{BH}$[M$_{\odot}$) = 8.03 $\pm$ 0.26).
Although the two distributions overlap, the NLS1 galaxies
have M$_{BH}$ values that are lower than those of the BLS1 galaxies. From the 
KS test statistics ($p$ = 7.9 $\times$ 10$^{-18}$), we found that the 
pressure-corrected M$_{BH}$ distribution of the BLS1 and NLS1 galaxies
is indeed different.

When we only consider the virial M$_{BH}$ values in NLS1 and BLS1 galaxies, 
the NLS1 galaxies in our sample have a mean M$_{BH}$ value of 
 log(M$_{BH}$[M$_{\odot}$]) = 6.86 $\pm$ 0.25, while the BLS1 galaxies have a mean value of log(M$_{BH}$[M$_{\odot}$]) = 7.98 $\pm$ 0.25.  
NLS1 galaxies are found to be powered by AGN with black hole masses that are almost an order of magnitude 
lower than BLS1 galaxies.
In contrast, after taking into account the effect of radiation pressure from 
ionising photons, the difference in the 
M$_{BH}$ values between BLS1 and NLS1 galaxies was reduced.
However, NLS1 
galaxies still have lower M$_{BH}$ values than BLS1 galaxies.
For NLS1 galaxies, we found a pressure-corrected mean 
log (M$_{BH}$[M$_{\odot}$]) of 7.45 $\pm$ 0.27 M$_{\odot}$, while for BLS1 
galaxies, we found a radiation-pressure-corrected mean 
log (M$_{BH}$[M$_{\odot}$]) = 8.04 $\pm$ 0.26 M$_{\odot}$. This means that the increase in BH mass of NLS1 after correction for radiation pressure is about 0.65 dex, or a factor of 5.
This agrees with the results of \cite{2008ApJ...678..693M}, who
found that when radiation pressure is taken into account, the BH masses of NLS1 are higher by a factor of 5.

\begin{figure}
\vbox{
\includegraphics[width=1.0\columnwidth]{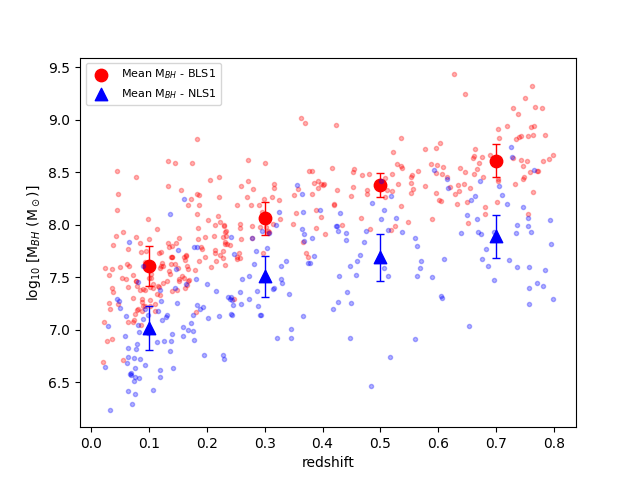}
\includegraphics[width=1.0\columnwidth]{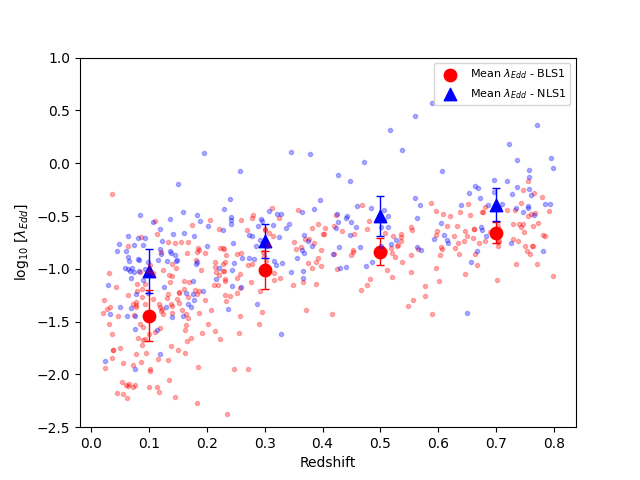} 
    }
\caption{Variation in the redshift-binned M$_{BH}$ (top) 
and $\lambda_{Edd}$ (bottom) for BLS1 (red) and NLS1 (blue) galaxies.
}
\label{figure-7}
\end{figure}

Using the radiation-pressure-corrected mean M$_{BH}$ values,
 we estimated the Eddington ratio $\lambda_{Edd} = $L$_{AGN}$ / L$_{Edd}$, 
where L$_{Edd} = 1.3 \times 10^{38} \times$ M$_{BH}$ / M$_{\odot}  (erg s^{-1}) $.
We found that NLS1 galaxies
have a higher Eddington ratio of log$_{10}$ ($\lambda_{Edd}$)
= $-$0.72 $\pm$ 0.22
than BLS1 galaxies, which have a mean 
log$_{10}$ ($\lambda_{Edd}$) = $-$1.08 $\pm$ 0.24.
The NLS1 galaxies have a higher Eddington ratio than BLS1 galaxies, which is
supported by a KS test statistics with a $p$-value of 2.48 $\times$ 10$^{-10}$. 
When binned in redshift, at each $z$ bin, NLS1 galaxies have lower M$_{BH}$ 
values than NLS1 galaxies (top panel of Fig. \ref{figure-7}).  In each 
$z$ bin, NLS1 galaxies also have higher $\lambda_{Edd}$ than BLS1 galaxies 
(Fig. \ref{figure-7} bottom panel).

\subsection{Correlation between star formation rate and active galactic nucleus luminosity}

\begin{figure*}
\includegraphics[width=1.0\columnwidth]{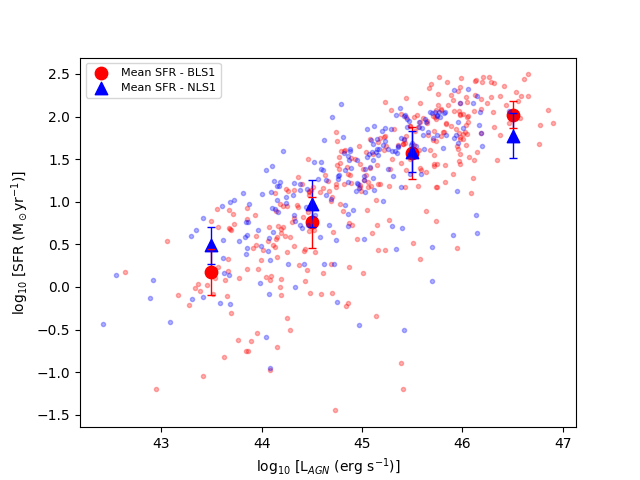}
\includegraphics[width=1.0\columnwidth]{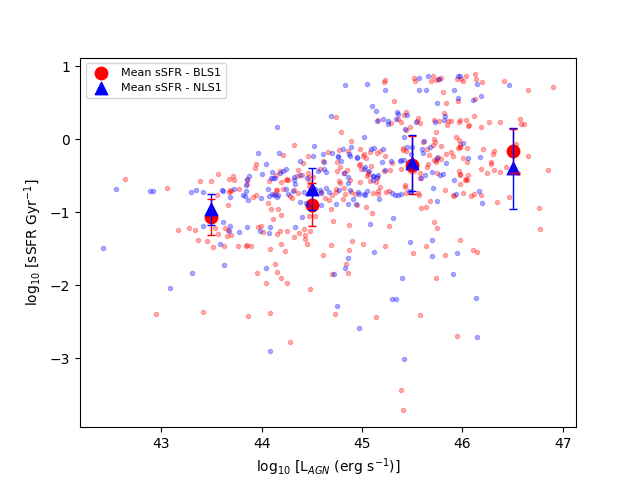}
\caption {Binned SFR (left) and sSFR (right) against L$_{AGN}$ for 
NLS1 (blue) and BLS1 (red) galaxies.}
\label{figure-8}
\end{figure*}

Previous studies of the correlation between SFR and
L$_{AGN}$ are inconclusive. A weak or
absent correlation was reported \citep{2015MNRAS.453..591S,2017MNRAS.472.2221S}, but also a strong 
correlation \citep{2017A&A...602A.123L,2017ApJ...835...74I,2020ApJ...896..108Z,2018A&A...618A..31M,2021A&A...646A.167M}. 
According to \cite{2015MNRAS.453..591S}, the observed flat relation may 
be due to short-timescale luminosity variation (driven by the accretion rate) in AGN, 
which may have washed out any inherent relation between SFR and L$_{AGN}$. Recently, 
\cite{2022ApJ...928...73K} found that AGN with a higher Eddington ratio and stronger 
outflows are hosted by galaxies with a high SFR, which is suggestive of no AGN 
feedback in quenching SF.
These conradicting results
might be due to the different methods that were used to estimate SFR and L$_{AGN}$.
We show in the left panel of Fig. \ref{figure-8} the correlation of SFR against L$_{AGN}$. Our
results show a strong positive correlation between
SFR and L$_{AGN}$, with a slope of 0.7 and 0.6 for BLS1 and NLS1 galaxies, respectively.
This is close to the slope value of 0.8 determined by \cite{2009MNRAS.399.1907N}.

This positive correlation could be an artefact due to the effects of M$_{\ast}$
and $z$ \citep{2017MNRAS.472.2221S}.  
To overcome the effect of M$_{\ast}$, we compared the sSFR with AGN
luminosities and found a much weaker correlation between the two properties.
The slightly high sSFRs at high AGN luminosities is a redshift effect, which 
can be seen by comparing Fig. \ref{figure-4} and Fig. \ref{figure-5}.
Thus, taking both M$_{\ast}$ and $z$ into consideration, we found no significant 
correlation between the sSFR and L$_{AGN}$.  Our results seem to indicate no 
impact of AGN feedback on both populations of Seyfert galaxies, in 
agreement with \cite{2017MNRAS.472.2221S}, who found a flat relation of SFR 
with L$_{AGN}$ for AGN with a range of luminosities. We note that the flat relation between SFR and AGN luminosity is attributed to the different timescales of the SF and AGN activity \citep{2014ApJ...782....9H}.
However, the spread in the sSFRs of AGN host 
galaxies is much larger than the 1$\sigma$ scatter in MS galaxies, which is 
around 0.3 dex \citep{2015A&A...575A..74S}.  This large scatter in the sSFRs of 
AGN host galaxies with respect to MS galaxies might be due to the impact of 
AGN on their host galaxies \citep{2018MNRAS.475.1288S}.

\subsection{Active galactic nucleus versus starbursts: Which dominates?}

\begin{figure*}
\includegraphics[width=1.0\columnwidth]{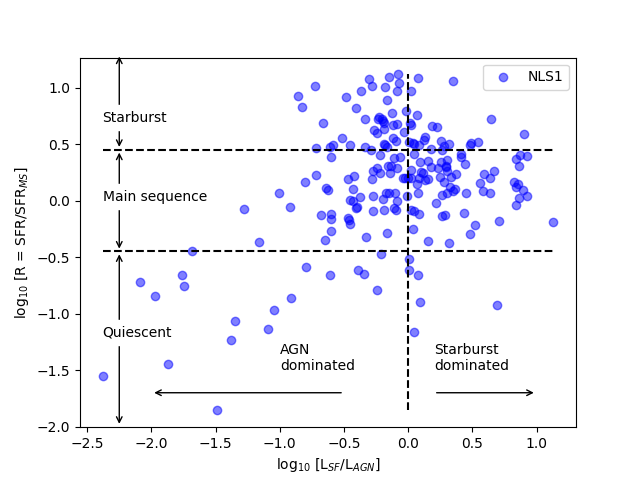}
\includegraphics[width=1.0\columnwidth]{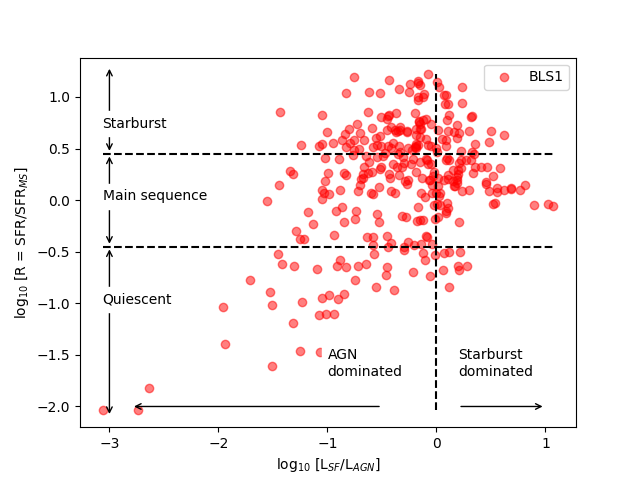}
\caption{Ratio of SFR to the main-sequence SFR of the corresponding stellar 
mass and redshift, R = SFR/SFR${_{MS}}$, vs. the ratio of SF 
luminosity to AGN luminosity, L${_{SF}}$/L${_{AGN}}$.}
\label{figure-9}
\end{figure*}

We show in Fig. \ref{figure-9} the location of our sample of sources in the R versus
log (L$_{SF}$/L$_{AGN}$) plane. Here, R is defined as the ratio of
SFR to the SFR$_{MS}$ for the corresponding M$_{\ast}$ and 
$z$, where SFR$_{MS}$ is estimated using Eq. 3. L$_{SF}$ is defined  
as the integrated luminosity due to the SF between 8 and 1000 
$\mu$m in its host galaxy. 
It is 
calculated from the SFR using the relation of \cite{1998ApJ...498..541K} corrected for the Chabrier IMF \citep{2010MNRAS.402.1693H},

\begin{equation}
\frac{L_{SF}}{L_{\odot}}  =   \frac{SFR}{1.1 \times 10^{-10}} \, erg s^{-1}.
\end{equation}

\cite{2015MNRAS.452.1841S} considered R to be the distance of a source from the SF MS. They considered objects with log(R) $>$ 1$\sigma$ to have enhanced SF. The 1$\sigma$
scatter in MS galaxies is 0.3 dex \citep{2015A&A...575A..74S}.
Conservatively, we considered a stricter limit of 1.5$\sigma$ and defined log(R) $>$ 0.45 dex as galaxies with enhanced SF,
$-$0.45 $<$ log(R) $<$ 0.45 as MS galaxies, and log(R) $<$ $-$0.45 dex as galaxies with suppressed SF.
In the sample of BLS1 galaxies, 35\% are above the MS, 48\% are on the MS, and 17\% are below the MS, whereas of the NLS1 galaxies, 31\% are above the MS, 57\% are on the MS, and 12\% are below the MS. The R-value estimated using the expressions in the literature may introduce systematic effects. However, in this paper, these systematics do not affect the overall results and conclusions because we compare the R between BLS1 and NLS1, and thus, these systematics would affect both population measurements in the same way. 

From \cite{2009MNRAS.399.1907N}, we define
a boundary in log(L$_{SF}$/L$_{AGN}$) wherein sources with log(L$_{SF}$/L$_{AGN}$)
$>$ 0 occupy the region dominated by starbursts and
sources with log(L$_{SF}$/L$_{AGN}$) $<$ 0 occupy the region dominated by
AGN.

We found that 71$\%$ of the BLS1 galaxies are AGN dominated, but only 55$\%$ of the NLS1 galaxies are AGN dominated.
This also indicates that NLS1 are more strongly dominated by starbursts (45$\%$) than BLS1 galaxies (29$\%$). While this is true, 
we recall that 
the SFR and sSFR properties are similar for both NLS1 and BLS1, as described in Section 4.1. 
Thus, that NLS1 galaxies are more strongly dominated by starbursts than BLS1 galaxies is 
largely driven by the differences in the AGN luminosities and not by their 
SF properties.

\subsection{Impact of active galactic nucleus activity on star formation}

\begin{figure*}
\includegraphics[width=1.0\columnwidth]{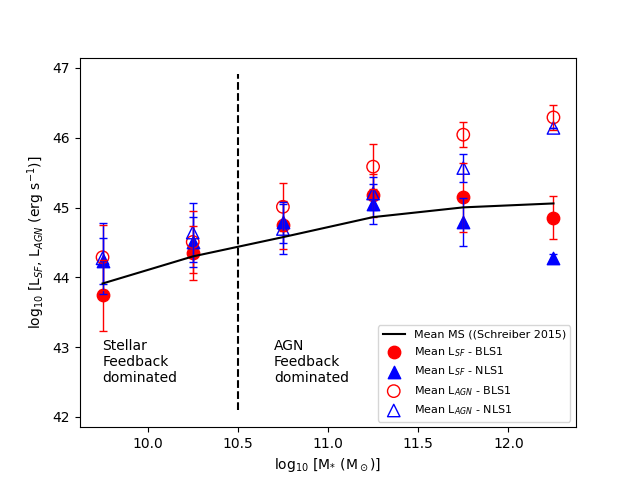}
\includegraphics[width=1.0\columnwidth]{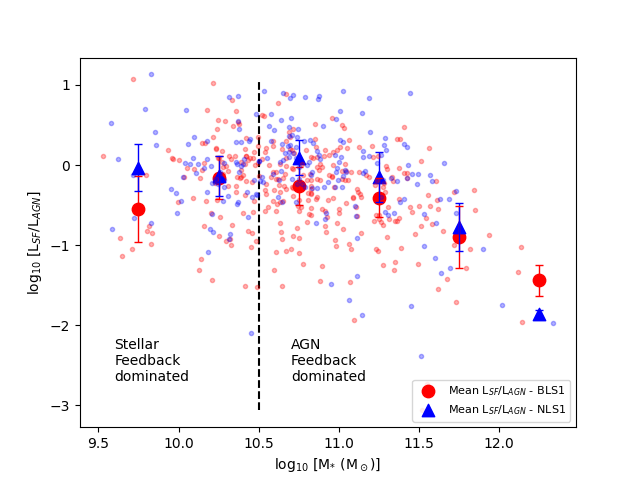}
\caption{Star formation luminosity (L${_{SF}}$) and AGN luminosity (L${_{AGN}}$) 
vs. stellar mass M$_{\ast}$ (left). Ratio of L${_{SF}}$ to L${_{AGN}}$ 
vs. stellar mass. (right)}
\label{figure-10}
\end{figure*}

\begin{figure}
\includegraphics[width=1.0\columnwidth]{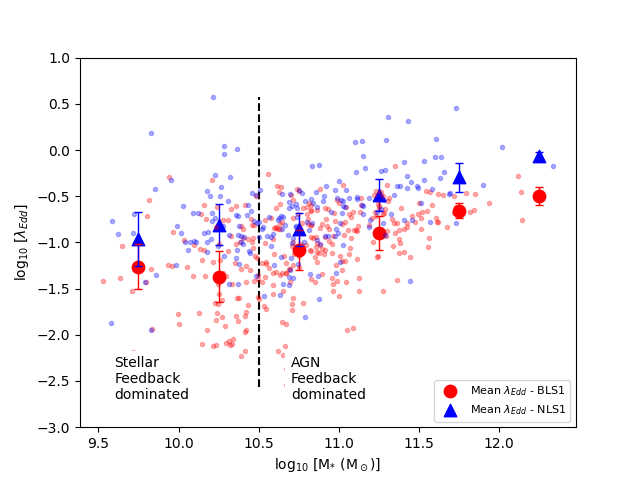}
\caption{Eddington ratio ($\lambda{_{Edd}}$) vs. stellar mass (M$_{\ast}$). }
\label{figure-11}
\end{figure}

Observations
are inconclusive about the role of AGN feedback in quenching SF and
its connection to the host galaxy evolution.
However, simulations of galaxy evolution are found to overpredict the high-mass
end of the stellar mass function, and to account for this 
quenching of SF, the AGN feedback process is
invoked \citep{2012MNRAS.422.2816B}.  Observationally, the number density of galaxies
per logarithmic mass bin is found to have a decreasing trend with mass, 
with a sharp cut-off at $\sim$10$^{10.5}$ M$_{\odot}$ \citep{2012MNRAS.421..621B,2015MNRAS.450.1937C}.
Towards the low-mass end, supernova feedback is thought to play
a role in regulating SF, while at the high-mass end,
feedback from AGN is invoked to play a role \citep{2015MNRAS.450.1937C}.
In high-redshift quasar hosts, a spatial anti-correlation is observed
between the wind component of [OIII] and  H$\alpha$ that traces the SF in the host galaxy \citep{2012A&A...537L...8C}. This could be 
evidence of quasar winds quenching SF within their disks \citep{2023hxga.book..126C}.

To investigate {the role of feedback, if any, in our sample of BLS1 and NLS1 galaxies} ,
we compare their L$_{SF}$ and L$_{AGN}$ luminosities with stellar 
mass in the left panel of Fig. \ref{figure-10}.
We also mark the change in the feedback modes from stellar feedback to AGN feedback for 
the transition mass of 3 $\times$ 10$^{10}$ M${_\odot}$, as known in 
the literature (eg. \citealt{2012MNRAS.421..621B}) and through simulations
(e.g. \citealt{2015MNRAS.450.1937C}).
For all further studies, we only considered the regime that is dominated by AGN feedback.

For both NLS1 and BLS1, the L$_{SF}$ initially increases with stellar mass, but then flattens at high stellar masses, while L$_{AGN}$ monotonically increases with M$_{\ast}$ up to the high stellar mass bin. In the left panel of Fig. \ref{figure-10}, we also plot the mean log (L$_{SF}$) of MS galaxies (solid black line), using Equation 5 for comparison. The L$_{SF}$ of NLS1 and BLS1 galaxies closely follow the relation of MS galaxies up to 
M$_{\ast}$ $\sim$ 10$^{11.5}$, beyond which it deviates. For further clarity, in the right panel of Fig. \ref{figure-10}, we show the change in 
L$_{SF}$/L$_{AGN}$ ratio with respect to stellar mass. 
\cite{2012ApJ...753L..30M} found that for stellar masses lower than 10$^{11}$ $M_{\odot}$, the L$_{SF}$/L$_{AGN}$ ratio remains constant, which they attributed to the availability of a common gas reservoir that regulates the AGN activity and SF. Our results agree with \cite{2012ApJ...753L..30M} for the same range of stellar masses.

From FIRE-3\footnote{Feedback In Realistic Environments (FIRE) 
project website: http://fire.northewestern.edu}  cosmological simulations with haloes in the mass
range of 10$^{12}$ $-$ 10$^{13}$ M$_{\odot}$, galaxies with AGN feedback
are found to have quenched SF. Alternatively, galaxies without AGN
are found to actively form stars regardless of their mass \citep{2023arXiv231016086B}.
These simulations also agree with observations 
\citep{2010ApJ...721..193P,2013ApJ...777...18M,2014ApJ...783...85T}, which indicate 
a reduced SF in galaxies with a stellar mass higher than 10$^{10.5}$ M$_{\odot}$, which 
corresponds to a halo mass of $\sim$10$^{12}$ M$_{\odot}$.

We found that L$_{SF}$/L$_{AGN}$ decreases with stellar mass. {\bf One of the reasons for this decrease might be} 
negative AGN feedback at higher stellar masses (see the right panel of Fig. \ref{figure-10}). 
This result is in line with 
simulations (e.g. \citealt{2018MNRAS.475.1288S,2023arXiv231016086B}). To verify the decreasing trend in L$_{SF}$/L$_{AGN}$,  we estimated the Spearman correlation coefficient for objects with a stellar mass greater than 10$^{10.5}$  M$_{\odot}$. The correlation coefficient and p-value for the BLS1 L$_{SF}$/L$_{AGN}$ and stellar mass are -0.30 and 3.8e-6, respectively. The correlation coefficient and p-value for the NLS1, L$_{SF}$/L$_{AGN}$, and stellar mass is -0.43 and 1.1e-7, respectively. These values suggest a weak negative correlation between L$_{SF}$/L$_{AGN}$ and  M$_{*}$ for the high stellar mass range. Our results from the analysis of BLS1 and NLS1 galaxies therefore agree with
simulations and with other observations in the literature.
In Appendix \ref{figure-14}, we also show that the redshift effects on this correlation are negligible. 

We also explore the change in Eddington ratio with stellar mass in 
Fig. \ref{figure-11}.
For the regime dominated by AGN feedback, the Eddington ratio increases with stellar mass, indicating an increasing radiative feedback with stellar mass, with more radiative feedback expected in NLS1 galaxies than BLS1 galaxies. Similar to the analysis done for L$_{SF}$/L$_{AGN}$, we verified the increasing trend by estimating the Spearman correlation coefficient for the Eddington ratio and M$_{*}$. The correlation coefficient and p-value for the BLS1 Eddington ratio and stellar mass are 0.40 and 6.1e-10, respectively. The correlation coefficient and p-value for the NLS1 Eddington ratio and stellar mass are 0.58 and 4.5e-14, respectively. These values suggest a weak positive correlation between the Eddington ratio and stellar mass for the high stellar mass range for BLS1 and a relatively strong positive correlation for NLS1. For AGN with high Eddington ratios, the radiative feedback from AGN can heat gas or remove gas from the host galaxies and suppress SF. This could be one reason for the decreasing trend in L$_{SF}$/L$_{AGN}$ with increasing stellar mass at high stellar masses for NLS1 when compared to BLS1 galaxies.

Based on simulations, \cite{2023arXiv231107576B} have found that 
radiative feedback from AGN can limit the gas inflow that powers
the SMBH and can also affect the host galaxy by suppressing the SF through
gas removal by AGN winds in the nuclear ($<$ kpc) region. This imprint
of AGN radiation is manifested at high accretion rates, wherein the gas
in the central nuclear region is swept away, leading to a quenching of the SF.
From spatially resolved observations of galaxies hosting AGN and normal
galaxies, \cite{2023ApJ...953...26L} found that in galaxies hosting AGN,
SFR is suppressed in the central region (kpc scale) compared to
normal galaxies. Although these observations recognise that AGN feedback
has an effect close to the central regions of their host galaxies, the
effect of AGN feedback affecting SF on the kiloparsec galaxy scale is uncertain 
\citep{2017A&A...601A.143F}.

An alternative scenario that can explain the decreasing trend in L$_{SF}$/L$_{AGN}$ or the flattening of L$_{SF}$ for high stellar masses is the decoupling of AGN fuelling from SF. \cite{2021A&A...653A..74M,2022A&A...663A.130M} proposed that in low stellar mass ranges ( $<$ 10$^{11}$M$_{\odot}$), the cold gas could fuel both the AGN and SF, whereas for high stellar mass ranges, the AGN-fuelling mechanism may be decoupled from SF. This scenario would also explain the positive trend seen in AGN and SF with respect to stellar mass for M$_{*}$ $<$ 10$^{11}$ M$_{\odot}$ in the left panel of Fig. \ref{figure-10}.

The MS galaxies are also known to flatten at 
high M$_{\ast}$ \citep{2015A&A...575A..74S,2019MNRAS.483.3213P}. One of the reasons for this flattening in MS galaxies according to 
semi-analytical and hydrodynamical
simulations might be the suppression of SF by SMBH feedback 
through outflows or jets that quench the bulge component and also prevent the 
cooling of gas in the disk component by feeding energy into the circumgalactic 
medium through jets and lobes  (see Section 4.1 of \cite{2019MNRAS.483.3213P}).
 Alternatively, the shape of the MS, including the flattening at high 
M$_{\ast}$, could be an outcome of the complex interplay between the morphology
of the galaxies and their environment, and might not solely be due to stellar mass
\cite{2019MNRAS.483.3213P}.

The BLS1 and NLS1 galaxies differ in their large-scale environment 
\citep{2017A&A...606A...9J}, and the 
galaxy environment may also be different  
among the individual sources analysed in this work. The environment can
have a strong impact on the AGN  activity as well as on SF, as
shown in simulations \citep{2023ApJ...953...64S, 2024A&A...683A..57R}. 
It is also likely that the nuclear
regions of BLS1 and NLS1 galaxies host complex mergers that are not seen in ground-based
optical/infrared images. This is supported by the identification of two galaxies
in the act of merging in an NLS1 galaxy using high-resolution adaptive optical 
J-band imaging \citep{2020ApJ...892..133P}. Spatially resolved observations of a large sample of sources are needed
to fully understand the effect of AGN feedback and the spatial scales at which they are prevalent.

The host morphology of BLS1 galaxies is likely to be
spiral galaxies \citep{1995ApJ...441...96M,1999ApJ...516..660H, 2011MNRAS.417.2721O}. Similarly,
NLS1 galaxies are also frequently found in disk systems 
\citep{2018A&A...619A..69J,2022A&A...668A..91V,2023A&A...679A..32V}. At low redshifts, \cite{2011MNRAS.417.2721O} found a median bulge-to-total ratio (B/T) of 0.39 in BLS1s and 0.17 in NLS1s and further stated that NLS1 have pseudo-bulges while BLS1 have a mix of classic and pseudo-bulges. However, it is also known that the bulge-to-total ratio increases with stellar mass (e.g. \citealt{2014ApJ...788...11L}). The growth of bulges in massive galaxies corresponds to a decline in gas mass fractions and SFRs (e.g. \citealt{2015ApJ...803...26P}). Star formation is linked to the disk component, which is the main driver of the MS relation. The growth of
the bulge increases the stellar mass of a galaxy, but does not alter its SFR \citep{2022MNRAS.513..256D}. It is therefore possible that due to the growth of the galaxy bulge at high stellar masses in Seyfert  1 galaxies, the contribution from the disk component towards SFR decreases, thus showing a flattened SFR. This may also decouple AGN fuelling from SF and would explain why the Eddington ratio increases with stellar mass while the SFR flattens or decreases.

\subsection{Effect of jets on star formation}
\begin{figure*}
  \hbox{
        \includegraphics[width=1.0\columnwidth]{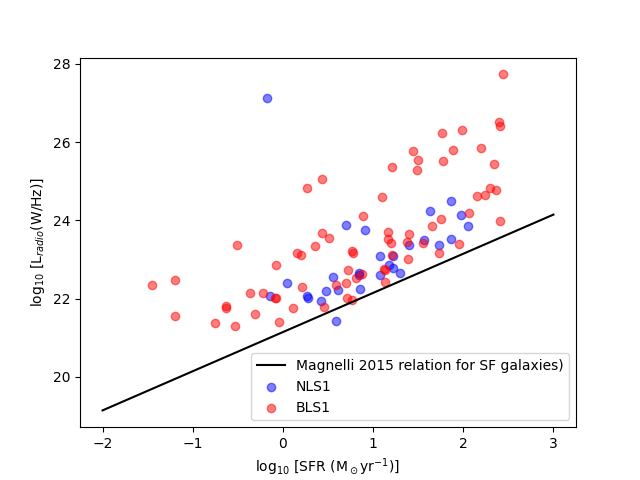}
        \includegraphics[width=1.0\columnwidth]{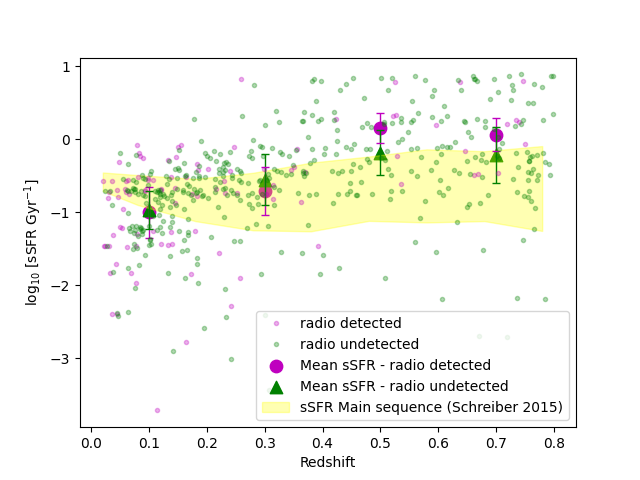}
       }
\caption{Comparison of radio-detected and radio-undetected Seyfert 1 galaxies. Left panel: Radio luminosity at 1.4 GHz  vs. the corresponding 
SFR. The black line shows the relation followed by star-forming galaxies. 
Right panel:  Location of the radio-detected and radio-undetected 
Seyfert 1 galaxies in the sSFR-z plane.}
\label{figure-12}
\end{figure*}

To investigate the effects of relativistic jets on the SF
characteristics of our sample of BLS1 and NLS1 galaxies, we
cross-correlated our sample with the sources in the Faint Images
of the Radio Sky at Twenty centimeters (FIRST; \citealt{1995ApJ...450..559B}) survey. We found 70 radio-detected BLS1 (or 21\% of the sample) and 29 radio-detected NLS1 (14\%). We show in
Fig. \ref{figure-12} the 1.4 GHz luminosity of the radio-detected sources against their
SFR. We also show in the same figure the relation
between the radio luminosity at 1.4 GHz and the SFR for star-forming 
galaxies \citep{2014A&A...561A..86M}. The FIRST-detected sources have a higher radio luminosity 
than expected for SF galaxies. This confirms the presence of 
relativistic jets in our radio-detected sample of BLS1 and NLS1 galaxies. 
We also show in Fig. \ref{figure-12} the variation in sSFR with $z$ for our 
sample of radio-detected and radio undetected sources compared with the 
sSFR of normal SF galaxies. We found that the sSFR of radio-detected 
and radio undetected sources in our sample is similar to the sSFR of normal
star-forming galaxies. We therefore conclude that the role of AGN jets, if any, in
altering the SF characteristics in our sample is negligible. However, we note that
available observational results in the literature indicate that radio jets affect the SF characteristics of their hosts \citep{2024ApJ...965...17D, 2023ApJ...959..116N}.

\section{Summary}\label{sec:summary}
We used the multi-band photometric data from the UV to the FIR to 
investigate the AGN and host galaxy properties of BLS1 and NLS1 galaxies. 
The findings of this work are summarised below.

\begin{enumerate}

\item The SFR and sSFR of NLS1 galaxies are similar to those of BLS1 
galaxies. The logarithm of the mean  SFR for BLS1 and NLS1  galaxies is
1.18 $\pm$ 0.41 and 1.18 $\pm$ 0.33  M$_{\odot}$/yr, respectively. Similarly,
the logarithm of the mean stellar masses is 10.78 $\pm$ 0.24 and 
10.74 $\pm$ 0.25 
M$_{\odot}$ for BLS1 and NLS1 galaxies, respectively. Thus, there is no 
difference in the SF properties and stellar masses in our sample of 
broad- and narrow-line Seyfert 1 galaxies. 

\item The logarithm of the mean sSFRs of NLS1 and BLS1 galaxies in each 
$z$ bin tends to lie on the SF main sequence in the sSFR-$z$ plane. 
This tends to indicate that the SF properties of Seyfert 1 galaxies are 
similar to the SF properties of MS galaxies.

\item The mean virial M$_{BH}$ estimates are similar to the mean radiation -pressure-corrected M$_{BH}$ values for BLS1 galaxies. However, 
for NLS1 galaxies, the virial M$_{BH}$ are underestimated. This 
is might be because the high radiation pressure dilutes the effects of gravity 
in NLS1 galaxies \citep{2008ApJ...678..693M}. 

\item NLS1 galaxies have a lower mean log (M$_{BH}$[M$_{\odot}$]) value of 
7.45 $\pm$ 0.27 than BLS1 galaxies, which have a mean 
log (M$_{BH}$ [M$_{\odot}$]) of
8.04 $\pm$ 0.26. This is true even when they are binned in redshift. This 
indicates that NLS1 are powered by lower-mass black holes than 
BLS1 galaxies.

\item BLS1 galaxies have more L$_{AGN}$ than { \bf NLS1} galaxies. 
We found a mean log(L$_{AGN}$) of 44.84 $\pm$ 0.39 erg s$^{-1}$ and
45.07 $\pm$ 0.44 erg s$^{-1}$ for NLS1 and BLS1 galaxies, respectively.

\item NLS1 have higher $\lambda_{Edd}$ than BLS1 galaxies. The reason might be 
that NLS1 galaxies are hosted in gas-rich sources with efficient 
black hole fuelling mechanisms such as bars \citep{2000MNRAS.314L..17M} and 
pseudo-bulges \citep{2012ApJ...754..146M}, compared to BLS1 galaxies.
For NLS1 galaxies, the mean logarithm of $\lambda_{Edd}$ is 
$-$0.72 $\pm$ 0.22, while for BLS1 galaxies, the mean logarithm
of $\lambda_{Edd}$ is $-$1.08 $\pm$ 0.24.
 
\item We found a strong positive correlation between SFR and  
L$_{AGN}$ for both BLS1 and NLS1 galaxies. When the effects of 
M$_{\ast}$ and $z$ were taken into account, no correlation was found. 

\item  
Using the log(L${_{SF}}$/L${_{AGN}}$) $>=$ 0 as an indicator of starburst-dominated sources, we found 
that 45$\%$ of the NLS1 are starburst dominated compared to 29 $\%$ of the BLS1 
galaxies. However, this is driven by the differences in the AGN luminosities 
of BLS1 and NLS1 galaxies.

\item
When comparing the log(L${_{SF}}$) and log(L${_{AGN}}$) ratio with stellar mass, we find
that as  log(L${_{AGN}}$) increases with stellar mass, log(L${_{SF}}$) flattens at high stellar masses
for both NLS1 and BLS1 galaxies.
This may be due to AGN activity, which might play a role in suppressing star 
formation at high stellar masses ($>$ 10$^{11}$ M$_{\odot}$) or to AGN fuelling, which might be decoupled from SF. However, the role
played by other factors such as environment and/or merging as well as the different
timescales of AGN and SF activity cannot be ruled out.  

\item When separated into radio-detected and radio-undetected sub-samples, 
we found no difference in the SF characteristics between these
two sub-samples. The relativistic jets in these sub-samples
have no impact on the SF properties of their
hosts. However, we note that our sample size is small, and 
a larger sample of radio-detected sources is required to study the 
effects of AGN jets on SF.
\end{enumerate}

This study indicates that NLS1 are low-mass black hole counterparts to BLS1 
galaxies with high Eddington ratios.  However, their SF properties are similar 
to each other and also to the SF of MS galaxies, but with large scatter.
Although SFR-L$_{AGN}$ shows a flat relation for both NLS1 and 
BLS1 galaxies, when we compared the ratio of SF luminosity with the AGN luminosity, we found 
that at high stellar masses, AGN feedback could play a role in suppressing SF, but there may be other physical processes as well. Spatially
resolved spectroscopic observations in the future hold the key to understanding the complex interplay between AGN activity and their hosts.

\begin{acknowledgements}
The funding provided by the Alexander von Humboldt Foundation, Germany is thankfully acknowledged.
K.S.K. acknowledges the ESO studentship provided by the European Southern Observatory, Garching.

\end{acknowledgements}

\bibliographystyle{aa} 
\bibliography{aanda_corr} 

\begin{thebibliography}{123}
\expandafter\ifx\csname natexlab\endcsname\relax\def\natexlab#1{#1}\fi

\bibitem[{{Abdo} {et~al.}(2009){Abdo}, {Ackermann}, {Ajello}, {Baldini},
  {Ballet}, {Barbiellini}, {Bastieri}, {Bechtol}, {Bellazzini}, {Berenji},
  {Bloom}, {Bonamente}, {Borgland}, {Bregeon}, {Brez}, {Brigida}, {Bruel},
  {Burnett}, {Caliandro}, {Cameron}, {Caraveo}, {Casandjian}, {Cecchi},
  {{\c{C}}elik}, {Chekhtman}, {Cheung}, {Chiang}, {Ciprini}, {Claus},
  {Cohen-Tanugi}, {Conrad}, {Cutini}, {Dermer}, {de Palma}, {Silva}, {Drell},
  {Dubois}, {Dumora}, {Farnier}, {Favuzzi}, {Fegan}, {Focke}, {Foschini},
  {Frailis}, {Fukazawa}, {Fusco}, {Gargano}, {Gehrels}, {Germani}, {Giebels},
  {Giglietto}, {Giordano}, {Giroletti}, {Glanzman}, {Godfrey}, {Grenier},
  {Grove}, {Guillemot}, {Guiriec}, {Hayashida}, {Hays}, {Horan}, {Hughes},
  {J{\'o}hannesson}, {Johnson}, {Johnson}, {Kadler}, {Kamae}, {Katagiri},
  {Kataoka}, {Kerr}, {Kn{\"o}dlseder}, {Kuss}, {Lande}, {Latronico}, {Longo},
  {Loparco}, {Lott}, {Lovellette}, {Lubrano}, {Makeev}, {Mazziotta},
  {McConville}, {McEnery}, {Meurer}, {Michelson}, {Mitthumsiri}, {Mizuno},
  {Monte}, {Monzani}, {Morselli}, {Moskalenko}, {Murgia}, {Nolan}, {Norris},
  {Nuss}, {Ohsugi}, {Omodei}, {Orlando}, {Ormes}, {Pelassa}, {Pepe}, {Persic},
  {Pesce-Rollins}, {Piron}, {Porter}, {Rain{\`o}}, {Rando}, {Razzano},
  {Rochester}, {Rodriguez}, {Ryde}, {Sadrozinski}, {Sambruna}, {Sander}, {Saz
  Parkinson}, {Scargle}, {Sgr{\`o}}, {Smith}, {Spandre}, {Spinelli},
  {Strickman}, {Suson}, {Tagliaferri}, {Takahashi}, {Takahashi}, {Tanaka},
  {Thayer}, {Thayer}, {Thompson}, {Tibaldo}, {Tibolla}, {Torres}, {Tosti},
  {Tramacere}, {Uchiyama}, {Usher}, {Vasileiou}, {Vilchez}, {Vitale}, {Waite},
  {Wang}, {Winer}, {Wood}, {Ylinen}, {Ziegler}, {Fermi/LAT Collaboration},
  {Ghisellini}, {Maraschi}, \& {Tavecchio}}]{2009ApJ...707L.142A}
{Abdo}, A.~A., {Ackermann}, M., {Ajello}, M., {et~al.} 2009, \apjl, 707, L142

\bibitem[{{Ant{\'o}n} {et~al.}(2008){Ant{\'o}n}, {Browne}, \&
  {March{\~a}}}]{2008A&A...490..583A}
{Ant{\'o}n}, S., {Browne}, I.~W.~A., \& {March{\~a}}, M.~J. 2008, \aap, 490,
  583

\bibitem[{{Babi{\'c}} {et~al.}(2007){Babi{\'c}}, {Miller}, {Jarvis}, {Turner},
  {Alexander}, \& {Croom}}]{2007A&A...474..755B}
{Babi{\'c}}, A., {Miller}, L., {Jarvis}, M.~J., {et~al.} 2007, \aap, 474, 755

\bibitem[{{Baes} {et~al.}(2011){Baes}, {Verstappen}, {De Looze}, {Fritz},
  {Saftly}, {Vidal P{\'e}rez}, {Stalevski}, \& {Valcke}}]{2011ApJS..196...22B}
{Baes}, M., {Verstappen}, J., {De Looze}, I., {et~al.} 2011, \apjs, 196, 22

\bibitem[{{Baldi} {et~al.}(2016){Baldi}, {Capetti}, {Robinson}, {Laor}, \&
  {Behar}}]{2016MNRAS.458L..69B}
{Baldi}, R.~D., {Capetti}, A., {Robinson}, A., {Laor}, A., \& {Behar}, E. 2016,
  \mnras, 458, L69

\bibitem[{{Baldry} {et~al.}(2012){Baldry}, {Driver}, {Loveday}, {Taylor},
  {Kelvin}, {Liske}, {Norberg}, {Robotham}, {Brough}, {Hopkins}, {Bamford},
  {Peacock}, {Bland-Hawthorn}, {Conselice}, {Croom}, {Jones}, {Parkinson},
  {Popescu}, {Prescott}, {Sharp}, \& {Tuffs}}]{2012MNRAS.421..621B}
{Baldry}, I.~K., {Driver}, S.~P., {Loveday}, J., {et~al.} 2012, \mnras, 421,
  621

\bibitem[{{Becker} {et~al.}(1995){Becker}, {White}, \&
  {Helfand}}]{1995ApJ...450..559B}
{Becker}, R.~H., {White}, R.~L., \& {Helfand}, D.~J. 1995, \apj, 450, 559

\bibitem[{{Bentz} {et~al.}(2013){Bentz}, {Denney}, {Grier}, {Barth},
  {Peterson}, {Vestergaard}, {Bennert}, {Canalizo}, {De Rosa}, {Filippenko},
  {Gates}, {Greene}, {Li}, {Malkan}, {Pogge}, {Stern}, {Treu}, \&
  {Woo}}]{2013ApJ...767..149B}
{Bentz}, M.~C., {Denney}, K.~D., {Grier}, C.~J., {et~al.} 2013, \apj, 767, 149

\bibitem[{{Bollati} {et~al.}(2023){Bollati}, {Lupi}, {Dotti}, \&
  {Haardt}}]{2023arXiv231107576B}
{Bollati}, F., {Lupi}, A., {Dotti}, M., \& {Haardt}, F. 2023, arXiv e-prints,
  arXiv:2311.07576

\bibitem[{{Boller} {et~al.}(1996){Boller}, {Brandt}, \&
  {Fink}}]{1996A&A...305...53B}
{Boller}, T., {Brandt}, W.~N., \& {Fink}, H. 1996, \aap, 305, 53

\bibitem[{{Bongiorno} {et~al.}(2012){Bongiorno}, {Merloni}, {Brusa},
  {Magnelli}, {Salvato}, {Mignoli}, {Zamorani}, {Fiore}, {Rosario}, {Mainieri},
  {Hao}, {Comastri}, {Vignali}, {Balestra}, {Bardelli}, {Berta}, {Civano},
  {Kampczyk}, {Le Floc'h}, {Lusso}, {Lutz}, {Pozzetti}, {Pozzi}, {Riguccini},
  {Shankar}, \& {Silverman}}]{2012MNRAS.427.3103B}
{Bongiorno}, A., {Merloni}, A., {Brusa}, M., {et~al.} 2012, \mnras, 427, 3103

\bibitem[{{Boquien} {et~al.}(2019){Boquien}, {Burgarella}, {Roehlly}, {Buat},
  {Ciesla}, {Corre}, {Inoue}, \& {Salas}}]{2019A&A...622A.103B}
{Boquien}, M., {Burgarella}, D., {Roehlly}, Y., {et~al.} 2019, \aap, 622, A103

\bibitem[{{Bower} {et~al.}(2012){Bower}, {Benson}, \&
  {Crain}}]{2012MNRAS.422.2816B}
{Bower}, R.~G., {Benson}, A.~J., \& {Crain}, R.~A. 2012, \mnras, 422, 2816

\bibitem[{{Bruzual} \& {Charlot}(2003)}]{2003MNRAS.344.1000B}
{Bruzual}, G. \& {Charlot}, S. 2003, \mnras, 344, 1000

\bibitem[{{Byrne} {et~al.}(2023){Byrne}, {Faucher-Gigu{\`e}re}, {Wellons},
  {Hopkins}, {Angl{\'e}s-Alc{\'a}zar}, {Sultan}, {Wijers}, {Moreno}, \&
  {Ponnada}}]{2023arXiv231016086B}
{Byrne}, L., {Faucher-Gigu{\`e}re}, C.-A., {Wellons}, S., {et~al.} 2023, arXiv
  e-prints, arXiv:2310.16086

\bibitem[{{Calderone} {et~al.}(2013){Calderone}, {Ghisellini}, {Colpi}, \&
  {Dotti}}]{2013MNRAS.431..210C}
{Calderone}, G., {Ghisellini}, G., {Colpi}, M., \& {Dotti}, M. 2013, \mnras,
  431, 210

\bibitem[{{Cano-D{\'\i}az} {et~al.}(2012){Cano-D{\'\i}az}, {Maiolino},
  {Marconi}, {Netzer}, {Shemmer}, \& {Cresci}}]{2012A&A...537L...8C}
{Cano-D{\'\i}az}, M., {Maiolino}, R., {Marconi}, A., {et~al.} 2012, \aap, 537,
  L8

\bibitem[{{Capelo} {et~al.}(2023){Capelo}, {Feruglio}, {Hickox}, \&
  {Tombesi}}]{2023hxga.book..126C}
{Capelo}, P.~R., {Feruglio}, C., {Hickox}, R.~C., \& {Tombesi}, F. 2023, in
  Handbook of X-ray and Gamma-ray Astrophysics. Edited by Cosimo Bambi and
  Andrea Santangelo, 126

\bibitem[{{Chabrier}(2003)}]{2003PASP..115..763C}
{Chabrier}, G. 2003, \pasp, 115, 763

\bibitem[{{Charlot} \& {Fall}(2000)}]{2000ApJ...539..718C}
{Charlot}, S. \& {Fall}, S.~M. 2000, \apj, 539, 718

\bibitem[{{Ciesla} {et~al.}(2015){Ciesla}, {Charmandaris}, {Georgakakis},
  {Bernhard}, {Mitchell}, {Buat}, {Elbaz}, {LeFloc'h}, {Lacey}, {Magdis}, \&
  {Xilouris}}]{2015A&A...576A..10C}
{Ciesla}, L., {Charmandaris}, V., {Georgakakis}, A., {et~al.} 2015, \aap, 576,
  A10

\bibitem[{{Ciesla} {et~al.}(2017){Ciesla}, {Elbaz}, \&
  {Fensch}}]{2017A&A...608A..41C}
{Ciesla}, L., {Elbaz}, D., \& {Fensch}, J. 2017, \aap, 608, A41

\bibitem[{{Crain} {et~al.}(2015){Crain}, {Schaye}, {Bower}, {Furlong},
  {Schaller}, {Theuns}, {Dalla Vecchia}, {Frenk}, {McCarthy}, {Helly},
  {Jenkins}, {Rosas-Guevara}, {White}, \& {Trayford}}]{2015MNRAS.450.1937C}
{Crain}, R.~A., {Schaye}, J., {Bower}, R.~G., {et~al.} 2015, \mnras, 450, 1937

\bibitem[{{Daddi} {et~al.}(2007){Daddi}, {Dickinson}, {Morrison}, {Chary},
  {Cimatti}, {Elbaz}, {Frayer}, {Renzini}, {Pope}, {Alexander}, {Bauer},
  {Giavalisco}, {Huynh}, {Kurk}, \& {Mignoli}}]{2007ApJ...670..156D}
{Daddi}, E., {Dickinson}, M., {Morrison}, G., {et~al.} 2007, \apj, 670, 156

\bibitem[{{Dale} {et~al.}(2014){Dale}, {Helou}, {Magdis}, {Armus},
  {D{\'\i}az-Santos}, \& {Shi}}]{2014ApJ...784...83D}
{Dale}, D.~A., {Helou}, G., {Magdis}, G.~E., {et~al.} 2014, \apj, 784, 83

\bibitem[{{Dickinson} {et~al.}(2003){Dickinson}, {Papovich}, {Ferguson}, \&
  {Budav{\'a}ri}}]{2003ApJ...587...25D}
{Dickinson}, M., {Papovich}, C., {Ferguson}, H.~C., \& {Budav{\'a}ri}, T. 2003,
  \apj, 587, 25

\bibitem[{{Dimauro} {et~al.}(2022){Dimauro}, {Daddi}, {Shankar}, {Cattaneo},
  {Huertas-Company}, {Bernardi}, {Caro}, {Dupke}, {H{\"a}u{\ss}ler},
  {Johnston}, {Cortesi}, {Mei}, \& {Peletier}}]{2022MNRAS.513..256D}
{Dimauro}, P., {Daddi}, E., {Shankar}, F., {et~al.} 2022, \mnras, 513, 256

\bibitem[{{Doi} {et~al.}(2012){Doi}, {Nagira}, {Kawakatu}, {Kino}, {Nagai}, \&
  {Asada}}]{2012ApJ...760...41D}
{Doi}, A., {Nagira}, H., {Kawakatu}, N., {et~al.} 2012, \apj, 760, 41

\bibitem[{{Doi} {et~al.}(2019){Doi}, {Nakahara}, {Nakamura}, {Kino},
  {Kawakatu}, \& {Nagai}}]{2019MNRAS.487..640D}
{Doi}, A., {Nakahara}, S., {Nakamura}, M., {et~al.} 2019, \mnras, 487, 640

\bibitem[{{Duggal} {et~al.}(2024){Duggal}, {O'Dea}, {Baum}, {Labiano},
  {Tadhunter}, {Worrall}, {Morganti}, {Tremblay}, \&
  {Dicken}}]{2024ApJ...965...17D}
{Duggal}, C., {O'Dea}, C.~P., {Baum}, S.~A., {et~al.} 2024, \apj, 965, 17

\bibitem[{{Elbaz} {et~al.}(2007){Elbaz}, {Daddi}, {Le Borgne}, {Dickinson},
  {Alexander}, {Chary}, {Starck}, {Brandt}, {Kitzbichler}, {MacDonald},
  {Nonino}, {Popesso}, {Stern}, \& {Vanzella}}]{2007A&A...468...33E}
{Elbaz}, D., {Daddi}, E., {Le Borgne}, D., {et~al.} 2007, \aap, 468, 33

\bibitem[{{Ferrarese} \& {Merritt}(2000)}]{2000ApJ...539L...9F}
{Ferrarese}, L. \& {Merritt}, D. 2000, \apjl, 539, L9

\bibitem[{{Fiore} {et~al.}(2017){Fiore}, {Feruglio}, {Shankar}, {Bischetti},
  {Bongiorno}, {Brusa}, {Carniani}, {Cicone}, {Duras}, {Lamastra}, {Mainieri},
  {Marconi}, {Menci}, {Maiolino}, {Piconcelli}, {Vietri}, \&
  {Zappacosta}}]{2017A&A...601A.143F}
{Fiore}, F., {Feruglio}, C., {Shankar}, F., {et~al.} 2017, \aap, 601, A143

\bibitem[{{Gebhardt} {et~al.}(2000){Gebhardt}, {Bender}, {Bower}, {Dressler},
  {Faber}, {Filippenko}, {Green}, {Grillmair}, {Ho}, {Kormendy}, {Lauer},
  {Magorrian}, {Pinkney}, {Richstone}, \& {Tremaine}}]{2000ApJ...539L..13G}
{Gebhardt}, K., {Bender}, R., {Bower}, G., {et~al.} 2000, \apjl, 539, L13

\bibitem[{{Grupe} \& {Mathur}(2004)}]{2004ApJ...606L..41G}
{Grupe}, D. \& {Mathur}, S. 2004, \apjl, 606, L41

\bibitem[{{G{\"u}ltekin} {et~al.}(2009){G{\"u}ltekin}, {Richstone}, {Gebhardt},
  {Lauer}, {Tremaine}, {Aller}, {Bender}, {Dressler}, {Faber}, {Filippenko},
  {Green}, {Ho}, {Kormendy}, {Magorrian}, {Pinkney}, \&
  {Siopis}}]{2009ApJ...698..198G}
{G{\"u}ltekin}, K., {Richstone}, D.~O., {Gebhardt}, K., {et~al.} 2009, \apj,
  698, 198

\bibitem[{{Harrison} {et~al.}(2012){Harrison}, {Alexander}, {Mullaney},
  {Altieri}, {Coia}, {Charmandaris}, {Daddi}, {Dannerbauer}, {Dasyra}, {Del
  Moro}, {Dickinson}, {Hickox}, {Ivison}, {Kartaltepe}, {Le Floc'h}, {Leiton},
  {Magnelli}, {Popesso}, {Rovilos}, {Rosario}, \&
  {Swinbank}}]{2012ApJ...760L..15H}
{Harrison}, C.~M., {Alexander}, D.~M., {Mullaney}, J.~R., {et~al.} 2012, \apjl,
  760, L15

\bibitem[{{Hatziminaoglou} {et~al.}(2009){Hatziminaoglou}, {Fritz}, \&
  {Jarrett}}]{2009MNRAS.399.1206H}
{Hatziminaoglou}, E., {Fritz}, J., \& {Jarrett}, T.~H. 2009, \mnras, 399, 1206

\bibitem[{{Hickox} {et~al.}(2014){Hickox}, {Mullaney}, {Alexander}, {Chen},
  {Civano}, {Goulding}, \& {Hainline}}]{2014ApJ...782....9H}
{Hickox}, R.~C., {Mullaney}, J.~R., {Alexander}, D.~M., {et~al.} 2014, \apj,
  782, 9

\bibitem[{{Hopkins} {et~al.}(2007){Hopkins}, {Richards}, \&
  {Hernquist}}]{2007ApJ...654..731H}
{Hopkins}, P.~F., {Richards}, G.~T., \& {Hernquist}, L. 2007, \apj, 654, 731

\bibitem[{{Hopkins} {et~al.}(2010){Hopkins}, {Younger}, {Hayward}, {Narayanan},
  \& {Hernquist}}]{2010MNRAS.402.1693H}
{Hopkins}, P.~F., {Younger}, J.~D., {Hayward}, C.~C., {Narayanan}, D., \&
  {Hernquist}, L. 2010, \mnras, 402, 1693

\bibitem[{{Hunt} \& {Malkan}(1999)}]{1999ApJ...516..660H}
{Hunt}, L.~K. \& {Malkan}, M.~A. 1999, \apj, 516, 660

\bibitem[{{Ichikawa} {et~al.}(2017){Ichikawa}, {Ricci}, {Ueda}, {Matsuoka},
  {Toba}, {Kawamuro}, {Trakhtenbrot}, \& {Koss}}]{2017ApJ...835...74I}
{Ichikawa}, K., {Ricci}, C., {Ueda}, Y., {et~al.} 2017, \apj, 835, 74

\bibitem[{{J{\"a}rvel{\"a}} {et~al.}(2018){J{\"a}rvel{\"a}},
  {L{\"a}hteenm{\"a}ki}, \& {Berton}}]{2018A&A...619A..69J}
{J{\"a}rvel{\"a}}, E., {L{\"a}hteenm{\"a}ki}, A., \& {Berton}, M. 2018, \aap,
  619, A69

\bibitem[{{J{\"a}rvel{\"a}} {et~al.}(2017){J{\"a}rvel{\"a}},
  {L{\"a}hteenm{\"a}ki}, {Lietzen}, {Poudel}, {Hein{\"a}m{\"a}ki}, \&
  {Einasto}}]{2017A&A...606A...9J}
{J{\"a}rvel{\"a}}, E., {L{\"a}hteenm{\"a}ki}, A., {Lietzen}, H., {et~al.} 2017,
  \aap, 606, A9

\bibitem[{{Jarvis} {et~al.}(2019){Jarvis}, {Harrison}, {Thomson}, {Circosta},
  {Mainieri}, {Alexander}, {Edge}, {Lansbury}, {Molyneux}, \&
  {Mullaney}}]{2019MNRAS.485.2710J}
{Jarvis}, M.~E., {Harrison}, C.~M., {Thomson}, A.~P., {et~al.} 2019, \mnras,
  485, 2710

\bibitem[{{Jha} {et~al.}(2022){Jha}, {Chand}, {Ojha}, {Omar}, \&
  {Rastogi}}]{2022MNRAS.510.4379J}
{Jha}, V.~K., {Chand}, H., {Ojha}, V., {Omar}, A., \& {Rastogi}, S. 2022,
  \mnras, 510, 4379

\bibitem[{{Kennicutt}(1998)}]{1998ApJ...498..541K}
{Kennicutt}, Robert~C., J. 1998, \apj, 498, 541

\bibitem[{{Kim} {et~al.}(2022){Kim}, {Woo}, {Jadhav}, {Chung}, {Baek}, {Lee},
  {Shin}, {Hwang}, {Luo}, {Son}, {Kim}, \& {Woo}}]{2022ApJ...928...73K}
{Kim}, C., {Woo}, J.-H., {Jadhav}, Y., {et~al.} 2022, \apj, 928, 73

\bibitem[{{Klimek} {et~al.}(2004){Klimek}, {Gaskell}, \&
  {Hedrick}}]{2004ApJ...609...69K}
{Klimek}, E.~S., {Gaskell}, C.~M., \& {Hedrick}, C.~H. 2004, \apj, 609, 69

\bibitem[{{Kormendy} \& {Richstone}(1995)}]{1995ARA&A..33..581K}
{Kormendy}, J. \& {Richstone}, D. 1995, \araa, 33, 581

\bibitem[{{Kotilainen} {et~al.}(2016){Kotilainen}, {Le{\'o}n-Tavares},
  {Olgu{\'\i}n-Iglesias}, {Baes}, {An{\'o}rve}, {Chavushyan}, \&
  {Carrasco}}]{2016ApJ...832..157K}
{Kotilainen}, J.~K., {Le{\'o}n-Tavares}, J., {Olgu{\'\i}n-Iglesias}, A.,
  {et~al.} 2016, \apj, 832, 157

\bibitem[{{Koutoulidis} {et~al.}(2022){Koutoulidis}, {Mountrichas},
  {Georgantopoulos}, {Pouliasis}, \& {Plionis}}]{2022A&A...658A..35K}
{Koutoulidis}, L., {Mountrichas}, G., {Georgantopoulos}, I., {Pouliasis}, E.,
  \& {Plionis}, M. 2022, \aap, 658, A35

\bibitem[{{Kshama} {et~al.}(2017){Kshama}, {Paliya}, \&
  {Stalin}}]{2017MNRAS.466.2679K}
{Kshama}, S.~K., {Paliya}, V.~S., \& {Stalin}, C.~S. 2017, \mnras, 466, 2679

\bibitem[{{Lammers} {et~al.}(2023){Lammers}, {Iyer}, {Ibarra-Medel},
  {Pacifici}, {S{\'a}nchez}, {Tacchella}, \& {Woo}}]{2023ApJ...953...26L}
{Lammers}, C., {Iyer}, K.~G., {Ibarra-Medel}, H., {et~al.} 2023, \apj, 953, 26

\bibitem[{{Lang} {et~al.}(2014){Lang}, {Wuyts}, {Somerville}, {F{\"o}rster
  Schreiber}, {Genzel}, {Bell}, {Brammer}, {Dekel}, {Faber}, {Ferguson},
  {Grogin}, {Kocevski}, {Koekemoer}, {Lutz}, {McGrath}, {Momcheva}, {Nelson},
  {Primack}, {Rosario}, {Skelton}, {Tacconi}, {van Dokkum}, \&
  {Whitaker}}]{2014ApJ...788...11L}
{Lang}, P., {Wuyts}, S., {Somerville}, R.~S., {et~al.} 2014, \apj, 788, 11

\bibitem[{{Lanzuisi} {et~al.}(2017){Lanzuisi}, {Delvecchio}, {Berta}, {Brusa},
  {Comastri}, {Gilli}, {Gruppioni}, {Marchesi}, {Perna}, {Pozzi}, {Salvato},
  {Symeonidis}, {Vignali}, {Vito}, {Volonteri}, \&
  {Zamorani}}]{2017A&A...602A.123L}
{Lanzuisi}, G., {Delvecchio}, I., {Berta}, S., {et~al.} 2017, \aap, 602, A123

\bibitem[{{Le{\'o}n Tavares} {et~al.}(2014){Le{\'o}n Tavares}, {Kotilainen},
  {Chavushyan}, {A{\~n}orve}, {Puerari}, {Cruz-Gonz{\'a}lez},
  {Pati{\~n}o-Alvarez}, {Ant{\'o}n}, {Carrami{\~n}ana}, {Carrasco}, {Guichard},
  {Karhunen}, {Olgu{\'\i}n-Iglesias}, {Sanghvi}, \&
  {Valdes}}]{2014ApJ...795...58L}
{Le{\'o}n Tavares}, J., {Kotilainen}, J., {Chavushyan}, V., {et~al.} 2014,
  \apj, 795, 58

\bibitem[{{Madau} \& {Dickinson}(2014)}]{2014ARA&A..52..415M}
{Madau}, P. \& {Dickinson}, M. 2014, \araa, 52, 415

\bibitem[{{Magnelli} {et~al.}(2014){Magnelli}, {Lutz}, {Saintonge}, {Berta},
  {Santini}, {Symeonidis}, {Altieri}, {Andreani}, {Aussel}, {B{\'e}thermin},
  {Bock}, {Bongiovanni}, {Cepa}, {Cimatti}, {Conley}, {Daddi}, {Elbaz},
  {F{\"o}rster Schreiber}, {Genzel}, {Ivison}, {Le Floc'h}, {Magdis},
  {Maiolino}, {Nordon}, {Oliver}, {Page}, {P{\'e}rez Garc{\'\i}a}, {Poglitsch},
  {Popesso}, {Pozzi}, {Riguccini}, {Rodighiero}, {Rosario}, {Roseboom},
  {Sanchez-Portal}, {Scott}, {Sturm}, {Tacconi}, {Valtchanov}, {Wang}, \&
  {Wuyts}}]{2014A&A...561A..86M}
{Magnelli}, B., {Lutz}, D., {Saintonge}, A., {et~al.} 2014, \aap, 561, A86

\bibitem[{{Magorrian} {et~al.}(1998){Magorrian}, {Tremaine}, {Richstone},
  {Bender}, {Bower}, {Dressler}, {Faber}, {Gebhardt}, {Green}, {Grillmair},
  {Kormendy}, \& {Lauer}}]{1998AJ....115.2285M}
{Magorrian}, J., {Tremaine}, S., {Richstone}, D., {et~al.} 1998, \aj, 115, 2285

\bibitem[{{Marconi} {et~al.}(2008){Marconi}, {Axon}, {Maiolino}, {Nagao},
  {Pastorini}, {Pietrini}, {Robinson}, \& {Torricelli}}]{2008ApJ...678..693M}
{Marconi}, A., {Axon}, D.~J., {Maiolino}, R., {et~al.} 2008, \apj, 678, 693

\bibitem[{{Marconi} \& {Hunt}(2003)}]{2003ApJ...589L..21M}
{Marconi}, A. \& {Hunt}, L.~K. 2003, \apjl, 589, L21

\bibitem[{{Masoura} {et~al.}(2021){Masoura}, {Mountrichas}, {Georgantopoulos},
  \& {Plionis}}]{2021A&A...646A.167M}
{Masoura}, V.~A., {Mountrichas}, G., {Georgantopoulos}, I., \& {Plionis}, M.
  2021, \aap, 646, A167

\bibitem[{{Masoura} {et~al.}(2018){Masoura}, {Mountrichas}, {Georgantopoulos},
  {Ruiz}, {Magdis}, \& {Plionis}}]{2018A&A...618A..31M}
{Masoura}, V.~A., {Mountrichas}, G., {Georgantopoulos}, I., {et~al.} 2018,
  \aap, 618, A31

\bibitem[{{Mathur}(2000)}]{2000MNRAS.314L..17M}
{Mathur}, S. 2000, \mnras, 314, L17

\bibitem[{{Mathur} {et~al.}(2012){Mathur}, {Fields}, {Peterson}, \&
  {Grupe}}]{2012ApJ...754..146M}
{Mathur}, S., {Fields}, D., {Peterson}, B.~M., \& {Grupe}, D. 2012, \apj, 754,
  146

\bibitem[{{McLeod} \& {Rieke}(1995)}]{1995ApJ...441...96M}
{McLeod}, K.~K. \& {Rieke}, G.~H. 1995, \apj, 441, 96

\bibitem[{{McLure} \& {Dunlop}(2002)}]{2002MNRAS.331..795M}
{McLure}, R.~J. \& {Dunlop}, J.~S. 2002, \mnras, 331, 795

\bibitem[{{Merritt} \& {Ferrarese}(2001)}]{2001ApJ...547..140M}
{Merritt}, D. \& {Ferrarese}, L. 2001, \apj, 547, 140

\bibitem[{{Morrissey} {et~al.}(2007){Morrissey}, {Conrow}, {Barlow}, {Small},
  {Seibert}, {Wyder}, {Budav{\'a}ri}, {Arnouts}, {Friedman}, {Forster},
  {Martin}, {Neff}, {Schiminovich}, {Bianchi}, {Donas}, {Heckman}, {Lee},
  {Madore}, {Milliard}, {Rich}, {Szalay}, {Welsh}, \&
  {Yi}}]{2007ApJS..173..682M}
{Morrissey}, P., {Conrow}, T., {Barlow}, T.~A., {et~al.} 2007, \apjs, 173, 682

\bibitem[{{Mountrichas} {et~al.}(2021{\natexlab{a}}){Mountrichas}, {Buat},
  {Georgantopoulos}, {Yang}, {Masoura}, {Boquien}, \&
  {Burgarella}}]{2021A&A...653A..70M}
{Mountrichas}, G., {Buat}, V., {Georgantopoulos}, I., {et~al.}
  2021{\natexlab{a}}, \aap, 653, A70

\bibitem[{{Mountrichas} {et~al.}(2021{\natexlab{b}}){Mountrichas}, {Buat},
  {Yang}, {Boquien}, {Burgarella}, {Ciesla}, {Malek}, \&
  {Shirley}}]{2021A&A...653A..74M}
{Mountrichas}, G., {Buat}, V., {Yang}, G., {et~al.} 2021{\natexlab{b}}, \aap,
  653, A74

\bibitem[{{Mountrichas} {et~al.}(2022){Mountrichas}, {Buat}, {Yang}, {Boquien},
  {Burgarella}, {Ciesla}, {Malek}, \& {Shirley}}]{2022A&A...663A.130M}
{Mountrichas}, G., {Buat}, V., {Yang}, G., {et~al.} 2022, \aap, 663, A130

\bibitem[{{Mullaney} {et~al.}(2015){Mullaney}, {Alexander}, {Aird}, {Bernhard},
  {Daddi}, {Del Moro}, {Dickinson}, {Elbaz}, {Harrison}, {Juneau}, {Liu},
  {Pannella}, {Rosario}, {Santini}, {Sargent}, {Schreiber}, {Simpson}, \&
  {Stanley}}]{2015MNRAS.453L..83M}
{Mullaney}, J.~R., {Alexander}, D.~M., {Aird}, J., {et~al.} 2015, \mnras, 453,
  L83

\bibitem[{{Mullaney} {et~al.}(2012){Mullaney}, {Daddi}, {B{\'e}thermin},
  {Elbaz}, {Juneau}, {Pannella}, {Sargent}, {Alexander}, \&
  {Hickox}}]{2012ApJ...753L..30M}
{Mullaney}, J.~R., {Daddi}, E., {B{\'e}thermin}, M., {et~al.} 2012, \apjl, 753,
  L30

\bibitem[{{Muzzin} {et~al.}(2013){Muzzin}, {Marchesini}, {Stefanon}, {Franx},
  {McCracken}, {Milvang-Jensen}, {Dunlop}, {Fynbo}, {Brammer}, {Labb{\'e}}, \&
  {van Dokkum}}]{2013ApJ...777...18M}
{Muzzin}, A., {Marchesini}, D., {Stefanon}, M., {et~al.} 2013, \apj, 777, 18

\bibitem[{{Nagao} {et~al.}(2002){Nagao}, {Murayama}, {Shioya}, \&
  {Taniguchi}}]{2002ApJ...575..721N}
{Nagao}, T., {Murayama}, T., {Shioya}, Y., \& {Taniguchi}, Y. 2002, \apj, 575,
  721

\bibitem[{{Nandi} {et~al.}(2023){Nandi}, {Stalin}, {Saikia}, {Riffel}, {Manna},
  {Pal}, {Dors}, {Wylezalek}, {Paliya}, {Saikia}, {Dabhade}, {Patig}, \&
  {Sagar}}]{2023ApJ...959..116N}
{Nandi}, P., {Stalin}, C.~S., {Saikia}, D.~J., {et~al.} 2023, \apj, 959, 116

\bibitem[{{Netzer}(2009)}]{2009MNRAS.399.1907N}
{Netzer}, H. 2009, \mnras, 399, 1907

\bibitem[{{Ojha} {et~al.}(2019){Ojha}, {Krishna}, \&
  {Chand}}]{2019MNRAS.483.3036O}
{Ojha}, V., {Krishna}, G., \& {Chand}, H. 2019, \mnras, 483, 3036

\bibitem[{{Olgu{\'\i}n-Iglesias} {et~al.}(2020){Olgu{\'\i}n-Iglesias},
  {Kotilainen}, \& {Chavushyan}}]{2020MNRAS.492.1450O}
{Olgu{\'\i}n-Iglesias}, A., {Kotilainen}, J., \& {Chavushyan}, V. 2020, \mnras,
  492, 1450

\bibitem[{{Orban de Xivry} {et~al.}(2011){Orban de Xivry}, {Davies},
  {Schartmann}, {Komossa}, {Marconi}, {Hicks}, {Engel}, \&
  {Tacconi}}]{2011MNRAS.417.2721O}
{Orban de Xivry}, G., {Davies}, R., {Schartmann}, M., {et~al.} 2011, \mnras,
  417, 2721

\bibitem[{{Osterbrock} \& {Pogge}(1985)}]{1985ApJ...297..166O}
{Osterbrock}, D.~E. \& {Pogge}, R.~W. 1985, \apj, 297, 166

\bibitem[{{Paliya} {et~al.}(2019){Paliya}, {Parker}, {Jiang}, {Fabian},
  {Brenneman}, {Ajello}, \& {Hartmann}}]{2019ApJ...872..169P}
{Paliya}, V.~S., {Parker}, M.~L., {Jiang}, J., {et~al.} 2019, \apj, 872, 169

\bibitem[{{Paliya} {et~al.}(2020){Paliya}, {P{\'e}rez}, {Garc{\'\i}a-Benito},
  {Ajello}, {Prada}, {Alberdi}, {Suh}, {Chandra}, {Dom{\'\i}nguez}, {Marchesi},
  {Di Matteo}, {Hartmann}, \& {Chiaberge}}]{2020ApJ...892..133P}
{Paliya}, V.~S., {P{\'e}rez}, E., {Garc{\'\i}a-Benito}, R., {et~al.} 2020,
  \apj, 892, 133

\bibitem[{{Paliya} {et~al.}(2013){Paliya}, {Stalin}, {Kumar}, {Kumar}, {Bhatt},
  {Pandey}, \& {Yadav}}]{2013MNRAS.428.2450P}
{Paliya}, V.~S., {Stalin}, C.~S., {Kumar}, B., {et~al.} 2013, \mnras, 428, 2450

\bibitem[{{Papovich} {et~al.}(2015){Papovich}, {Labb{\'e}}, {Quadri}, {Tilvi},
  {Behroozi}, {Bell}, {Glazebrook}, {Spitler}, {Straatman}, {Tran}, {Cowley},
  {Dav{\'e}}, {Dekel}, {Dickinson}, {Ferguson}, {Finkelstein}, {Gawiser},
  {Inami}, {Faber}, {Kacprzak}, {Kawinwanichakij}, {Kocevski}, {Koekemoer},
  {Koo}, {Kurczynski}, {Lotz}, {Lu}, {Lucas}, {McIntosh}, {Mehrtens},
  {Mobasher}, {Monson}, {Morrison}, {Nanayakkara}, {Persson}, {Salmon},
  {Simons}, {Tomczak}, {van Dokkum}, {Weiner}, \&
  {Willner}}]{2015ApJ...803...26P}
{Papovich}, C., {Labb{\'e}}, I., {Quadri}, R., {et~al.} 2015, \apj, 803, 26

\bibitem[{{Peng} {et~al.}(2010){Peng}, {Lilly}, {Kova{\v{c}}}, {Bolzonella},
  {Pozzetti}, {Renzini}, {Zamorani}, {Ilbert}, {Knobel}, {Iovino}, {Maier},
  {Cucciati}, {Tasca}, {Carollo}, {Silverman}, {Kampczyk}, {de Ravel},
  {Sanders}, {Scoville}, {Contini}, {Mainieri}, {Scodeggio}, {Kneib}, {Le
  F{\`e}vre}, {Bardelli}, {Bongiorno}, {Caputi}, {Coppa}, {de la Torre},
  {Franzetti}, {Garilli}, {Lamareille}, {Le Borgne}, {Le Brun}, {Mignoli},
  {Perez Montero}, {Pello}, {Ricciardelli}, {Tanaka}, {Tresse}, {Vergani},
  {Welikala}, {Zucca}, {Oesch}, {Abbas}, {Barnes}, {Bordoloi}, {Bottini},
  {Cappi}, {Cassata}, {Cimatti}, {Fumana}, {Hasinger}, {Koekemoer},
  {Leauthaud}, {Maccagni}, {Marinoni}, {McCracken}, {Memeo}, {Meneux}, {Nair},
  {Porciani}, {Presotto}, \& {Scaramella}}]{2010ApJ...721..193P}
{Peng}, Y.-j., {Lilly}, S.~J., {Kova{\v{c}}}, K., {et~al.} 2010, \apj, 721, 193

\bibitem[{{Pilbratt} {et~al.}(2010){Pilbratt}, {Riedinger}, {Passvogel},
  {Crone}, {Doyle}, {Gageur}, {Heras}, {Jewell}, {Metcalfe}, {Ott}, \&
  {Schmidt}}]{2010A&A...518L...1P}
{Pilbratt}, G.~L., {Riedinger}, J.~R., {Passvogel}, T., {et~al.} 2010, \aap,
  518, L1

\bibitem[{{Popesso} {et~al.}(2019){Popesso}, {Concas}, {Morselli}, {Schreiber},
  {Rodighiero}, {Cresci}, {Belli}, {Erfanianfar}, {Mancini}, {Inami},
  {Dickinson}, {Ilbert}, {Pannella}, \& {Elbaz}}]{2019MNRAS.483.3213P}
{Popesso}, P., {Concas}, A., {Morselli}, L., {et~al.} 2019, \mnras, 483, 3213

\bibitem[{{Rakshit} {et~al.}(2019){Rakshit}, {Johnson}, {Stalin}, {Gandhi}, \&
  {Hoenig}}]{2019MNRAS.483.2362R}
{Rakshit}, S., {Johnson}, A., {Stalin}, C.~S., {Gandhi}, P., \& {Hoenig}, S.
  2019, \mnras, 483, 2362

\bibitem[{{Rakshit} \& {Stalin}(2017)}]{2017ApJ...842...96R}
{Rakshit}, S. \& {Stalin}, C.~S. 2017, \apj, 842, 96

\bibitem[{{Rakshit} {et~al.}(2017){Rakshit}, {Stalin}, {Chand}, \&
  {Zhang}}]{2017ApJS..229...39R}
{Rakshit}, S., {Stalin}, C.~S., {Chand}, H., \& {Zhang}, X.-G. 2017, \apjs,
  229, 39

\bibitem[{{Rakshit} {et~al.}(2018){Rakshit}, {Stalin}, {Hota}, \&
  {Konar}}]{2018ApJ...869..173R}
{Rakshit}, S., {Stalin}, C.~S., {Hota}, A., \& {Konar}, C. 2018, \apj, 869, 173

\bibitem[{{Rani} {et~al.}(2017){Rani}, {Stalin}, \&
  {Rakshit}}]{2017MNRAS.466.3309R}
{Rani}, P., {Stalin}, C.~S., \& {Rakshit}, S. 2017, \mnras, 466, 3309

\bibitem[{{Rihtar{\v{s}}i{\v{c}}} {et~al.}(2024){Rihtar{\v{s}}i{\v{c}}},
  {Biffi}, {Fabjan}, \& {Dolag}}]{2024A&A...683A..57R}
{Rihtar{\v{s}}i{\v{c}}}, G., {Biffi}, V., {Fabjan}, D., \& {Dolag}, K. 2024,
  \aap, 683, A57

\bibitem[{{Sani} {et~al.}(2010){Sani}, {Lutz}, {Risaliti}, {Netzer}, {Gallo},
  {Trakhtenbrot}, {Sturm}, \& {Boller}}]{2010MNRAS.403.1246S}
{Sani}, E., {Lutz}, D., {Risaliti}, G., {et~al.} 2010, \mnras, 403, 1246

\bibitem[{{Santini} {et~al.}(2012){Santini}, {Rosario}, {Shao}, {Lutz},
  {Maiolino}, {Alexander}, {Altieri}, {Andreani}, {Aussel}, {Bauer}, {Berta},
  {Bongiovanni}, {Brandt}, {Brusa}, {Cepa}, {Cimatti}, {Daddi}, {Elbaz},
  {Fontana}, {F{\"o}rster Schreiber}, {Genzel}, {Grazian}, {Le Floc'h},
  {Magnelli}, {Mainieri}, {Nordon}, {P{\'e}rez Garcia}, {Poglitsch}, {Popesso},
  {Pozzi}, {Riguccini}, {Rodighiero}, {Salvato}, {Sanchez-Portal}, {Sturm},
  {Tacconi}, {Valtchanov}, \& {Wuyts}}]{2012A&A...540A.109S}
{Santini}, P., {Rosario}, D.~J., {Shao}, L., {et~al.} 2012, \aap, 540, A109

\bibitem[{{Scholtz} {et~al.}(2018){Scholtz}, {Alexander}, {Harrison},
  {Rosario}, {McAlpine}, {Mullaney}, {Stanley}, {Simpson}, {Theuns}, {Bower},
  {Hickox}, {Santini}, \& {Swinbank}}]{2018MNRAS.475.1288S}
{Scholtz}, J., {Alexander}, D.~M., {Harrison}, C.~M., {et~al.} 2018, \mnras,
  475, 1288

\bibitem[{{Schreiber} {et~al.}(2015){Schreiber}, {Pannella}, {Elbaz},
  {B{\'e}thermin}, {Inami}, {Dickinson}, {Magnelli}, {Wang}, {Aussel}, {Daddi},
  {Juneau}, {Shu}, {Sargent}, {Buat}, {Faber}, {Ferguson}, {Giavalisco},
  {Koekemoer}, {Magdis}, {Morrison}, {Papovich}, {Santini}, \&
  {Scott}}]{2015A&A...575A..74S}
{Schreiber}, C., {Pannella}, M., {Elbaz}, D., {et~al.} 2015, \aap, 575, A74

\bibitem[{{Shimizu} {et~al.}(2015){Shimizu}, {Mushotzky}, {Mel{\'e}ndez},
  {Koss}, \& {Rosario}}]{2015MNRAS.452.1841S}
{Shimizu}, T.~T., {Mushotzky}, R.~F., {Mel{\'e}ndez}, M., {Koss}, M., \&
  {Rosario}, D.~J. 2015, \mnras, 452, 1841

\bibitem[{{Shimizu} {et~al.}(2017){Shimizu}, {Mushotzky}, {Mel{\'e}ndez},
  {Koss}, {Barger}, \& {Cowie}}]{2017MNRAS.466.3161S}
{Shimizu}, T.~T., {Mushotzky}, R.~F., {Mel{\'e}ndez}, M., {et~al.} 2017,
  \mnras, 466, 3161

\bibitem[{{Silverman} {et~al.}(2009){Silverman}, {Lamareille}, {Maier},
  {Lilly}, {Mainieri}, {Brusa}, {Cappelluti}, {Hasinger}, {Zamorani},
  {Scodeggio}, {Bolzonella}, {Contini}, {Carollo}, {Jahnke}, {Kneib}, {Le
  F{\`e}vre}, {Merloni}, {Bardelli}, {Bongiorno}, {Brunner}, {Caputi},
  {Civano}, {Comastri}, {Coppa}, {Cucciati}, {de la Torre}, {de Ravel},
  {Elvis}, {Finoguenov}, {Fiore}, {Franzetti}, {Garilli}, {Gilli}, {Iovino},
  {Kampczyk}, {Knobel}, {Kova{\v{c}}}, {Le Borgne}, {Le Brun}, {Mignoli},
  {Pello}, {Peng}, {Perez Montero}, {Ricciardelli}, {Tanaka}, {Tasca},
  {Tresse}, {Vergani}, {Vignali}, {Zucca}, {Bottini}, {Cappi}, {Cassata},
  {Fumana}, {Griffiths}, {Kartaltepe}, {Koekemoer}, {Marinoni}, {McCracken},
  {Memeo}, {Meneux}, {Oesch}, {Porciani}, \& {Salvato}}]{2009ApJ...696..396S}
{Silverman}, J.~D., {Lamareille}, F., {Maier}, C., {et~al.} 2009, \apj, 696,
  396

\bibitem[{{Singh} {et~al.}(2023){Singh}, {Park}, {Choi}, {Kim}, {Jun},
  {Gibson}, {Kim}, {Lee}, \& {Snaith}}]{2023ApJ...953...64S}
{Singh}, A., {Park}, C., {Choi}, E., {et~al.} 2023, \apj, 953, 64

\bibitem[{{Skrutskie} {et~al.}(2006){Skrutskie}, {Cutri}, {Stiening},
  {Weinberg}, {Schneider}, {Carpenter}, {Beichman}, {Capps}, {Chester},
  {Elias}, {Huchra}, {Liebert}, {Lonsdale}, {Monet}, {Price}, {Seitzer},
  {Jarrett}, {Kirkpatrick}, {Gizis}, {Howard}, {Evans}, {Fowler}, {Fullmer},
  {Hurt}, {Light}, {Kopan}, {Marsh}, {McCallon}, {Tam}, {Van Dyk}, \&
  {Wheelock}}]{2006AJ....131.1163S}
{Skrutskie}, M.~F., {Cutri}, R.~M., {Stiening}, R., {et~al.} 2006, \aj, 131,
  1163

\bibitem[{{Stalevski} {et~al.}(2016){Stalevski}, {Ricci}, {Ueda}, {Lira},
  {Fritz}, \& {Baes}}]{2016MNRAS.458.2288S}
{Stalevski}, M., {Ricci}, C., {Ueda}, Y., {et~al.} 2016, \mnras, 458, 2288

\bibitem[{{Stanley} {et~al.}(2017){Stanley}, {Alexander}, {Harrison},
  {Rosario}, {Wang}, {Aird}, {Bourne}, {Dunne}, {Dye}, {Eales}, {Knudsen},
  {Micha{\l}owski}, {Valiante}, {De Zotti}, {Furlanetto}, {Ivison}, {Maddox},
  \& {Smith}}]{2017MNRAS.472.2221S}
{Stanley}, F., {Alexander}, D.~M., {Harrison}, C.~M., {et~al.} 2017, \mnras,
  472, 2221

\bibitem[{{Stanley} {et~al.}(2018){Stanley}, {Harrison}, {Alexander},
  {Simpson}, {Knudsen}, {Mullaney}, {Rosario}, \&
  {Scholtz}}]{2018MNRAS.478.3721S}
{Stanley}, F., {Harrison}, C.~M., {Alexander}, D.~M., {et~al.} 2018, \mnras,
  478, 3721

\bibitem[{{Stanley} {et~al.}(2015){Stanley}, {Harrison}, {Alexander},
  {Swinbank}, {Aird}, {Del Moro}, {Hickox}, \&
  {Mullaney}}]{2015MNRAS.453..591S}
{Stanley}, F., {Harrison}, C.~M., {Alexander}, D.~M., {et~al.} 2015, \mnras,
  453, 591

\bibitem[{{Tomczak} {et~al.}(2014){Tomczak}, {Quadri}, {Tran}, {Labb{\'e}},
  {Straatman}, {Papovich}, {Glazebrook}, {Allen}, {Brammer}, {Kacprzak},
  {Kawinwanichakij}, {Kelson}, {McCarthy}, {Mehrtens}, {Monson}, {Persson},
  {Spitler}, {Tilvi}, \& {van Dokkum}}]{2014ApJ...783...85T}
{Tomczak}, A.~R., {Quadri}, R.~F., {Tran}, K.-V.~H., {et~al.} 2014, \apj, 783,
  85

\bibitem[{{Turner} {et~al.}(2022){Turner}, {Miller}, {Maune}, \&
  {Eggen}}]{2022MNRAS.517.3257T}
{Turner}, C.~S., {Miller}, H.~R., {Maune}, J.~D., \& {Eggen}, J.~R. 2022,
  \mnras, 517, 3257

\bibitem[{{Varglund} {et~al.}(2023){Varglund}, {J{\"a}rvel{\"a}}, {Ciroi},
  {Berton}, {Congiu}, {L{\"a}hteenm{\"a}ki}, \& {Di
  Mille}}]{2023A&A...679A..32V}
{Varglund}, I., {J{\"a}rvel{\"a}}, E., {Ciroi}, S., {et~al.} 2023, \aap, 679,
  A32

\bibitem[{{Varglund} {et~al.}(2022){Varglund}, {J{\"a}rvel{\"a}},
  {L{\"a}hteenm{\"a}ki}, {Berton}, {Ciroi}, \& {Congiu}}]{2022A&A...668A..91V}
{Varglund}, I., {J{\"a}rvel{\"a}}, E., {L{\"a}hteenm{\"a}ki}, A., {et~al.}
  2022, \aap, 668, A91

\bibitem[{{Vietri} {et~al.}(2022{\natexlab{a}}){Vietri}, {J{\"a}rvel{\"a}},
  {Berton}, {Ciroi}, {Congiu}, {Chen}, \& {Di Mille}}]{2022A&A...662A..20V}
{Vietri}, A., {J{\"a}rvel{\"a}}, E., {Berton}, M., {et~al.} 2022{\natexlab{a}},
  \aap, 662, A20

\bibitem[{{Vietri} {et~al.}(2022{\natexlab{b}}){Vietri}, {Garilli}, {Polletta},
  {Bisogni}, {Cassar{\`a}}, {Franzetti}, {Fumana}, {Gargiulo}, {Maccagni},
  {Mancini}, {Scodeggio}, {Fritz}, {Ma{\l}ek}, {Manzoni}, {Pollo}, {Siudek},
  {Vergani}, {Zamorani}, \& {Zanichelli}}]{2022A&A...659A.129V}
{Vietri}, G., {Garilli}, B., {Polletta}, M., {et~al.} 2022{\natexlab{b}}, \aap,
  659, A129

\bibitem[{{Viswanath} {et~al.}(2019){Viswanath}, {Stalin}, {Rakshit}, {Kurian},
  {Ujjwal}, {Gudennavar}, \& {Kartha}}]{2019ApJ...881L..24V}
{Viswanath}, G., {Stalin}, C.~S., {Rakshit}, S., {et~al.} 2019, \apjl, 881, L24

\bibitem[{{Whitaker} {et~al.}(2014){Whitaker}, {Franx}, {Leja}, {van Dokkum},
  {Henry}, {Skelton}, {Fumagalli}, {Momcheva}, {Brammer}, {Labb{\'e}},
  {Nelson}, \& {Rigby}}]{2014ApJ...795..104W}
{Whitaker}, K.~E., {Franx}, M., {Leja}, J., {et~al.} 2014, \apj, 795, 104

\bibitem[{{Williams} {et~al.}(2018){Williams}, {Gliozzi}, \&
  {Rudzinsky}}]{2018MNRAS.480...96W}
{Williams}, J.~K., {Gliozzi}, M., \& {Rudzinsky}, R.~V. 2018, \mnras, 480, 96

\bibitem[{{Wright} {et~al.}(2010){Wright}, {Eisenhardt}, {Mainzer}, {Ressler},
  {Cutri}, {Jarrett}, {Kirkpatrick}, {Padgett}, {McMillan}, {Skrutskie},
  {Stanford}, {Cohen}, {Walker}, {Mather}, {Leisawitz}, {Gautier}, {McLean},
  {Benford}, {Lonsdale}, {Blain}, {Mendez}, {Irace}, {Duval}, {Liu}, {Royer},
  {Heinrichsen}, {Howard}, {Shannon}, {Kendall}, {Walsh}, {Larsen}, {Cardon},
  {Schick}, {Schwalm}, {Abid}, {Fabinsky}, {Naes}, \&
  {Tsai}}]{2010AJ....140.1868W}
{Wright}, E.~L., {Eisenhardt}, P. R.~M., {Mainzer}, A.~K., {et~al.} 2010, \aj,
  140, 1868

\bibitem[{{Zhou} {et~al.}(2007){Zhou}, {Wang}, {Yuan}, {Shan}, {Komossa}, {Lu},
  {Liu}, {Xu}, {Bai}, \& {Jiang}}]{2007ApJ...658L..13Z}
{Zhou}, H., {Wang}, T., {Yuan}, W., {et~al.} 2007, \apjl, 658, L13

\bibitem[{{Zhuang} \& {Ho}(2020)}]{2020ApJ...896..108Z}
{Zhuang}, M.-Y. \& {Ho}, L.~C. 2020, \apj, 896, 108

\bibitem[{{Zhuang} \& {Ho}(2022)}]{2022ApJ...934..130Z}
{Zhuang}, M.-Y. \& {Ho}, L.~C. 2022, \apj, 934, 130

\end{thebibliography}

\begin{appendix} 

\section{Effect of not including FIR data in the spectral energy distribution modelling}

\begin{figure}
\vbox{
\includegraphics[scale=0.55]{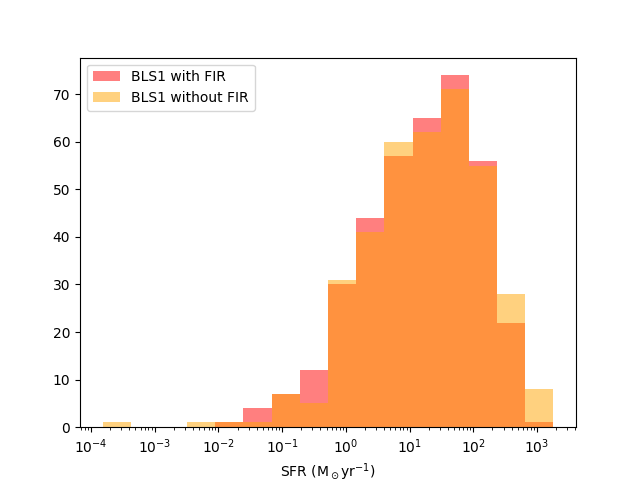}
\includegraphics[scale=0.55]{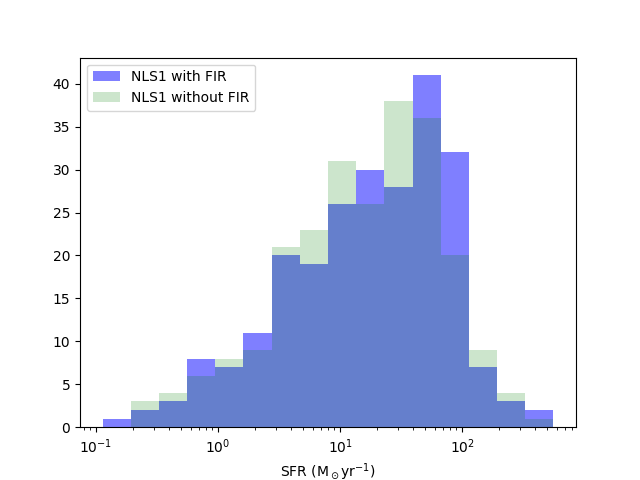}
     }
\caption{Distribution of the SFR with and without the FIR for BLS1 galaxies (top 
panel) and NLS1 galaxies (bottom panel).}
\label{figure-13}
\end{figure}

Our current sample of BLS1 and NLS1 galaxies was made with the need for 
a FIR detection for robust SFR estimates. It has been 
pointed out by \cite{2015A&A...576A..10C} that not including or a lack of FIR 
data points in the broad-band SED analysis cannot provide a reliable estimate 
of the SFR. To test this, we performed two sets of SED analyses, one set with FIR points, and the other set without the FIR points. We 
investigated the obtained physical parameters with and without the 
FIR photometric points in 
the SED modelling. We found that when the FIR points are not included, the derived physical parameters are not affected. We show in 
Fig. \ref{figure-13} the distribution of the SFR with and without the 
FIR for the sample of the BLS1 and NLS1 galaxies. The distribution is 
indistinguishable. We conclude that the non-inclusion of the FIR does not affect the 
parameters derived in this work.  

\section{Effect of redshift on L${_{SF}}$/L${_{AGN}}$ 
versus stellar mass }

\begin{figure}
\vbox{
\includegraphics[scale=0.55]{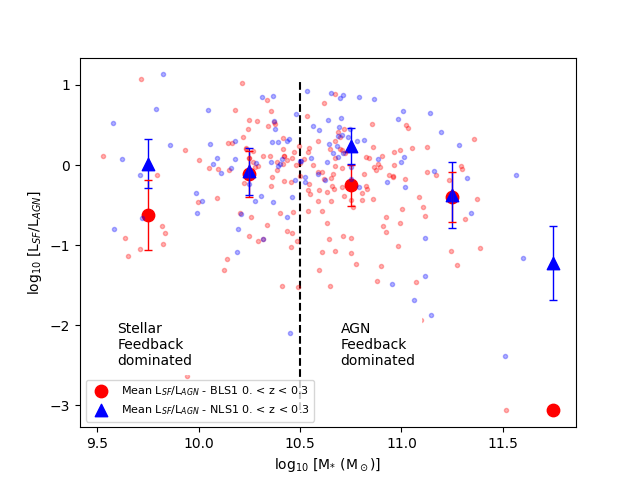}
\includegraphics[scale=0.55]{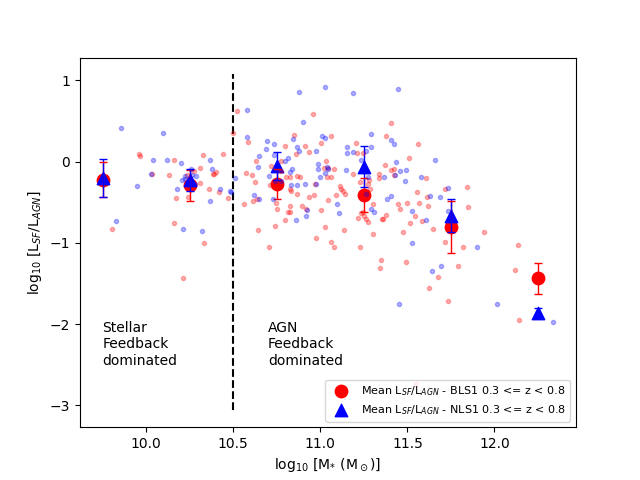}
     }
\caption{Ratio of L${_{SF}}$ to L${_{AGN}}$ 
vs. stellar mass for BLS1 galaxies and NLS1 galaxies in the low-redshift (top panel) and high-redshift regions (bottom panel).}
\label{figure-14}
\end{figure}

In Fig. \ref{figure-14}  we show the variation of L${_{SF}}$/L${_{AGN}}$ for the low-redshift ($0. < z < 0.3$) and high-redshift ($0.3 <= z < 0.8$) ranges. For both redshift bins, a decreasing value of L${_{SF}}$/L${_{AGN}}$ with increasing stellar mass is observed, but with a large scatter in the low-redshift range. { At low redshifts, the BLS1s L$_{SF}$/L$_{AGN}$ and stellar mass have a correlation co-efficient of -0.14 and a p-value of 0.16, and NLS1s have a correlation coefficient of -0.34 and a p-value of 1.2e-3. The BLS1 low-redshift sample shows no correlation between L$_{SF}$/L$_{AGN}$ and M$_{*}$. However, the BLS1 low-redshift sample is statistically poor in the highest M$_{*}$ bin. For the high-redshift range, the correlation coefficient and p-value for the BLS1 L$_{SF}$/L$_{AGN}$ and stellar mass are -0.37 and 2.7e-5, respectively, and the correlation coefficient and p-value for the NLS1s at high redshifts is -0.37 and 6.4e-4, respectively. These values suggest a weak negative correlation between L$_{SF}$/L$_{AGN}$ and  M$_{*}$ for the high stellar mass range, similar to the values for the full redshift range (Section 4.5 and right panel of Fig. \ref{figure-10}). The NLS1 sample at high and low redshifts shows similar correlation coefficients. We therefore conclude that the effect of the redshift on the ratio of L${_{SF}}$ to L${_{AGN}}$ is negligible for the NLS1 sample.}

\section{Comparing the parent sample and the Herschel FIR-selected sample}

\begin{figure*}
  \hbox{
        \includegraphics[width=1.0\columnwidth]{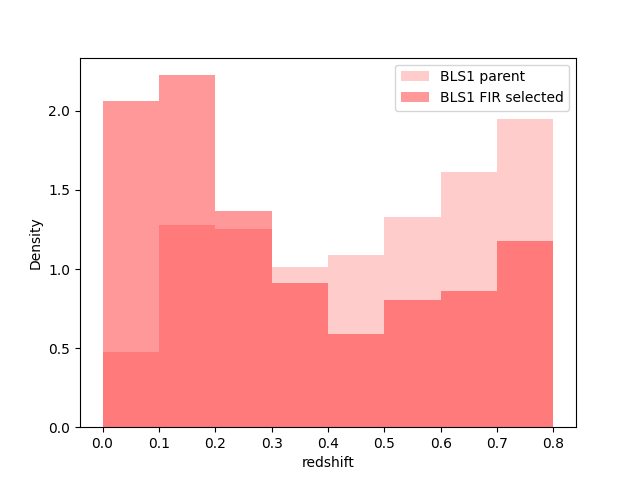}
        \includegraphics[width=1.0\columnwidth]{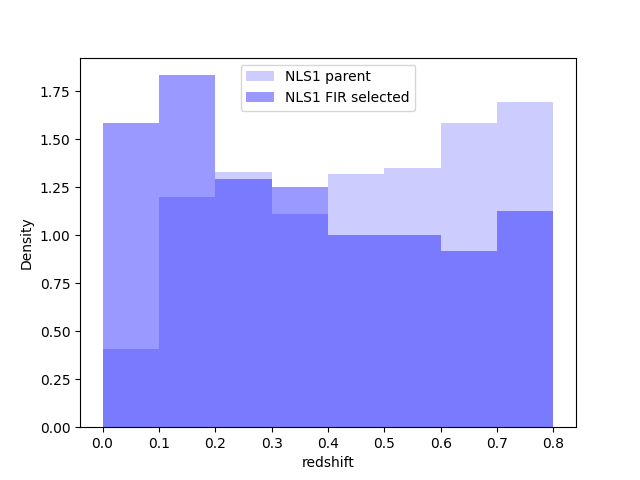}
       }
\caption{Distribution of z for the parent sample and FIR-selected sample for BLS1 galaxies (left 
panel) and NLS1 galaxies (right panel).}
\label{figure-15}
\end{figure*}

The SDSS-selected parent sample consists of 11101 and 14886 NLS1 and BLS1 galaxies. With
the Herschel FIR constraint described in Section 2, our sample for this study consisted of 240 NLS1 galaxies and 373 BLS1 galaxies. Fig.\ref{figure-15} shows that the FIR-selected sample does not have a similar distribution to the parent sample for the NLS1 and BLS1. The FIR selection biases the sample at high redshifts towards high SFR systems. However, this bias affects both NLS1 and BLS1 equally and does not affect the results of the comparative analysis.

\section{Fluxes}

The observed fluxes and their corresponding errors for our sample of BLS1 galaxies and NLS1 galaxies are given in the Tables D.1 and D.2 \footnote {Tables D.1 and D.2 are only available in electronic form at the CDS via anonymous ftp to cdsarc.u-strasbg.fr (130.79.128.5) or via http://cdsweb.u-strasbg.fr/cgi-bin/qcat?J/A+A/.}, respectively. We note that the selected sample of 240 NLS1 and 373 BLS1 galaxies was made for the broad-band SED modelling, but the final sample selected for the comparative study was based on the SED fitting (see Section 3). All the fluxes are in units of mJy. The columns are ordered in the following manner: Column 1 - index, Column - SDSS-ID, Column 3 \& 4- RA \& Dec, Column 5 - redshift z, Column 6 \& 7 - FUV flux \& FUV$_{err}$, Column 8 \& 9 - NUV flux \& NUV$_{err}$, Column 10 \& 11 - SDSS u' \& u'$_{err}$, Column 12 \& 13 - SDSS g' \& g'$_{err}$, Column 14 \& 15 - SDSS r' \& r'$_{err}$, Column 16 \& 17 - SDSS i' \& i'$_{err}$, Column 18 \& 19 - SDSS z' \& z'$_{err}$, Column 20 \& 21 - 2Mass J \& J$_{err}$, Column 22 \& 23 - 2Mass H \& H$_{err}$, Column 24 \& 25 - 2Mass Ks \& Ks$_{err}$, Column 26 \& 27 - WISE W1 \& W1$_{err}$, Column 28 \& 29 - WISE W2 \& W2$_{err}$, Column 30 \& 31 - WISE W3 \& W3$_{err}$, Column 32 \& 33 - WISE W4 \& W4$_{err}$, Column 34 \& 35 - Herschel PACS (green) \& PACS$_{err}$ (green), Column 36 \& 37 - Herschel PACS (red) \& PACS$_{err}$ (red), Column 38 \& 39 - Herschel SPIRE PSW \& PSW$_{err}$, Column 40 \& 41 - Herschel SPIRE PMW \& PMW$_{err}$, Column 42 \& 43 - Herschel SPIRE PLW \& PLW$_{err}$.








\end{appendix}

\end{document}